\documentclass[10pt,journal,compsoc]{IEEEtran}

\usepackage{hyperref}
\hypersetup{colorlinks=true,urlcolor=black,linkcolor=black,citecolor=black}
\usepackage{adjustbox} 
\usepackage{booktabs} 
\usepackage{multirow} 
\usepackage{microtype}
\usepackage{relsize}
\usepackage{color,xcolor}
\usepackage{subcaption}
\usepackage{comment}

\usepackage[all]{nowidow}

\usepackage{lipsum}
\usepackage[switch,columnwise]{lineno}

\usepackage{listings} 
\definecolor{light-gray}{gray}{0.95}

\lstMakeShortInline[columns=fixed]|

\lstdefinelanguage{LOGIC}{ 
 identifierstyle=\color{black},
keywordstyle=\bfseries,
keywords={implies,always,if,return,double,event},
otherkeywords={&,<,>,*,:,U,=},
morecomment=[l][\color{blue}]{//}
}
\lstdefinelanguage{RML}{ 
 stringstyle=\color{mauve},
 identifierstyle=\color{black},
keywordstyle=\bfseries,
keywords={position,matches,Main,pose,with, topic,>>,||},
morecomment=[l][\color{blue}]{//},
showstringspaces=false
}

\lstdefinelanguage{Python}{ 
 stringstyle=\color{mauve},
 identifierstyle=\color{black},
keywordstyle=\bfseries,
keywords={def,for, if, break, in}
}  

\lstdefinelanguage{Cpp}{ 
 stringstyle=\color{mauve},
 identifierstyle=\color{black},
keywordstyle=\bfseries,
keywords={void,for,if, new, bool, string},
morecomment=[l][\color{blue}]{//}
}

\lstdefinelanguage{BT}{ 
 stringstyle=\color{mauve},
 identifierstyle=\color{black},
keywordstyle=\bfseries,
keywords={behaviortree, condition, maneuver, subtree,?,->},
morecomment=[l][\color{blue}]{!--},
showstringspaces=false
}

\lstdefinelanguage{XML} { 
 stringstyle=\color{mauve},
 identifierstyle=\color{black},
keywordstyle=\bfseries,
keywords={xml, version, encoding, UTF, id, obstacle, waypoint, agent, source, scenario, addwaypoint},
morecomment=[l][\color{blue}]{!--},
showstringspaces=false
}      

\lstdefinelanguage{SQL} { 
 identifierstyle=\color{black},
keywordstyle=\bfseries,
keywords={manipulator,ros__parameters,modes,__DEFAULT__,max_torque,WEAK,STRONG}
} 

\newcommand{\rvft}{runtime verification and field-based testing}

\usepackage{fontawesome} 

\usepackage[capitalize]{cleveref}
\crefname{section}{Sect.}{sections}
\Crefname{section}{Section}{Sections}

\begin{document}

\title{Runtime Verification and Field-based Testing for ROS-based Robotic Systems}

\author{
Ricardo Caldas, Juan Antonio Pi\~{n}era Garc\'{i}a,  Matei Schiopu,\\ Patrizio~Pelliccione, Genaína~Rodrigues, and Thorsten Berger
\thanks{
R.~Caldas and M. Schiopu are with Chalmers University of Technology, Gothenburg, Sweden, - e-mail: 
\{\{ricardo.caldas,schiopu\}@chalmers.se\}\newline
J. A. Pi\~{n}era Garc\'{i}a and P.~Pelliccione are with Gran Sasso Science Institute (GSSI), L'Aquila, Italy - e-mail: \{\{antonio.pinera,patrizio.pelliccione\}@gssi.it\}\newline
G. Rodrigues is with University of Brasilia, Brasília,  Brazil - e-mail:
\{genaina@unb.br\}\newline
T. Berger is with Ruhr University Bochum, Germany and Chalmers and 
University of Gothenburg, Sweden - e-mail: 
\{thorsten.berger@rub.de\}
}
}

\IEEEtitleabstractindextext{%
\begin{abstract}
Robotic systems are becoming pervasive and adopted in increasingly many domains, such as manufacturing, healthcare, and space exploration. To this end, engineering software has emerged as a crucial discipline for building maintainable and reusable robotic systems. The field of robotics software engineering research has received increasing attention, fostering autonomy as a fundamental goal. However, robotics developers are still challenged trying to achieve this goal given that simulation is not able to deliver solutions to realistically emulate real-world phenomena. Robots also need to operate in unpredictable and uncontrollable environments, which require safe and trustworthy self-adaptation capabilities implemented in software. Typical techniques to address the challenges are runtime verification, field-based testing, and mitigation techniques that enable fail-safe solutions. However, there is no clear guidance to architect ROS-based systems to enable and facilitate runtime verification and field-based testing. This paper aims to fill in this gap by providing guidelines that can help developers and quality assurance (QA) teams when developing, verifying or testing their robots in the field. These guidelines are carefully tailored to address the challenges and requirements of testing robotics systems in real-world scenarios. We conducted (i) a literature review on studies addressing runtime verification and field-based testing for robotic systems, (ii)  mined ROS-based applications repositories, and (iii) validated the applicability, clarity, and usefulness via two questionnaires with 55 answers overall. We contribute 20 guidelines: 8 for developers and 12 for QA teams formulated for researchers and practitioners in robotic software engineering. Finally, we map our guidelines to open challenges thus far in runtime verification and field-based testing for ROS-based systems and, we outline promising research directions in the field.

\noindent\textbf{Guidelines website and replication package:} \href{https://ros-rvft.github.io}{https://ros-rvft.github.io}
\end{abstract}

}

\makeatletter
\newcommand{\linebreakand}{%
  \end{@IEEEauthorhalign}
  \hfill\mbox{}\par
  \mbox{}\hfill\begin{@IEEEauthorhalign}
}

 \makeatletter
 \def\ps@IEEEtitlepagestyle{
         \def\@oddfoot{\mycopyrightnotice}
         \def\@evenfoot{}
 }
 \def\mycopyrightnotice{
         {\footnotesize
                 \begin{minipage}{\textwidth}
                         \centering
                         \textcopyright~2024 IEEE.  Personal use of this material is 
 permitted.  Permission from IEEE must be obtained for all other uses, in
 any current or future media, including reprinting/republishing this
 material for advertising or promotional purposes, creating new
 collective works, for resale or redistribution to servers or lists, or
 reuse of any copyrighted component of this work in other works.
                 \end{minipage}
         }
 }

\maketitle

\section{Introduction}

Robotics has become fundamental to societal advancement, featuring applications that range from micromachinery for medicine and healthcare to space exploration and navigation~\cite{scirobotics}. 
As robots become increasingly pervasive, the role of software in robotic systems is rising significantly~\cite{brugali2009software,garcia2020robotics}. For instance, considering service robotics, studies have shown that a significant amount of resources is spent on building software rather than hardware~\cite{blumlein2014function}. The importance of software can be explained by the need to deploy robots with high autonomy and self-adaptation capabilities, allowing robots to operate under uncertainty~\cite{camara2020software,Askarpour2021}. 

In the light of the growing recognition of software's pivotal role in robotics, software engineering is crucial for mission specification~\cite{dragule2021languages,menghi2022mission,menghi:2021,dragule.ea:2021:sosym}, system architecture definition~\cite{kortenkamp2016robotic,rodrigues2022architecture}, component design,  implementation~\cite{brugaliCBREpart1,brugaliCBREpart2,ghzouli2023behavior} as well as verification and validation~\cite{gotlieb2021testing, cavalcanti2021robostar, ingrand2021verification}. Achieving autonomy, which minimizes human intervention, is a fundamental goal in robotics. However, despite the progress made, robotic systems still face challenges in adapting to real-world scenarios~\cite{bozhinoski2019safety}. There is a need for instruments to deal with the openness and uncertainty of robots' operational environments -- requiring (self-) adaptiveness capabilities and specific quality assurance techniques. 

The toughest challenges in engineering software for robotics are: achieving robustness, costly and slow quality assurance provision processes, and lack of means to design dynamic (self-)adaptation~\cite{garcia2020robotics,EMSE2022}. 
Furthermore, current simulation solutions are not able to emulate real-world phenomena in a sufficiently realistic manner, so developers prefer real-world experimentation to simulation~\cite{garcia2020robotics,EMSE2022}. The robotics domain specificity is that (i) robots act in the real world, which introduces all sorts of unexpected error cases that should be caught and managed, (ii) the robots can change the environment itself via actions, and (iii) due to uncertainty and high variability, it is difficult to model real-world environments where robots operate, especially those that involve humans~\cite{garcia2020robotics,EMSE2022}. 

In this context, it is impractical or even impossible to have robot's correct behavior guarantees at design time. To collect guarantees in robotics, quality assurance techniques must anticipate faulty scenarios, in spite of the complexities of runtime. Complexities arise from large input spaces, unpredictable corner cases, and lack of an oracle. Typical techniques to mitigate such complexities include using heuristics (e.g., test coverage~\cite{hildebrandt2023physcov}), or exploiting information available only at runtime~\cite{rizwan2023ezskiros, hildebrandt2021world}). It is unclear, however, how to systematically adopt such techniques in a real-world setting.

ROS (Robot Operating System) is the de facto standard for robot application development and it has revolutionized robotics software engineering. However, there are still limited advances in quality assurance methods for ROS-based systems. Alami et. al.~\cite{alami2018} studied the effects of standards and practices of the ROS community contributing to quality assurance practices, highlighting what the influencers to successful adoption of quality assurance in the ROS community are. Moreover, tooling such as HAROS~\cite{santos:2021} promotes property-based testing of ROS applications by automatically deriving test cases from requirements specified in formal properties and the source code, which enables gaining confidence in ROS applications. However, there are no reported guidelines on available tooling and quality assurance techniques tackling real-world complexities for gaining confidence in ROS-based applications. 

Our goal is to address this gap by establishing guidelines that can help developers and quality assurance (QA) teams when developing, verifying (runtime verification), or testing their robots in the field. In fact, despite recent efforts in providing guidance to architecting ROS-based systems~\cite{malavolta2021mining}, there is no clear guidance on how to design ROS-based systems to enable and facilitate quality assurance with verification and testing, especially at runtime. Our guidelines are tailored to ROS-based systems. They are defined not only for QA teams but also for developers. QA teams are responsible for the robotic systems' verification and validation. Developers are responsible for preparing software for field-based testing and runtime verification. Our research questions 
are:

\vspace{1mm}

\noindent\textbf{RQ1.} What guidelines should robotics developers follow to design ROS-based systems in order to enable and facilitate runtime verification and field-based testing?

\vspace{1mm}

\noindent\textbf{RQ2.} What guidelines should quality assurance teams follow to perform runtime verification and field-based testing for ROS-based applications? 

\vspace{1mm}

\begin{figure}[t!]
    \centering
    \includegraphics[width=\columnwidth]{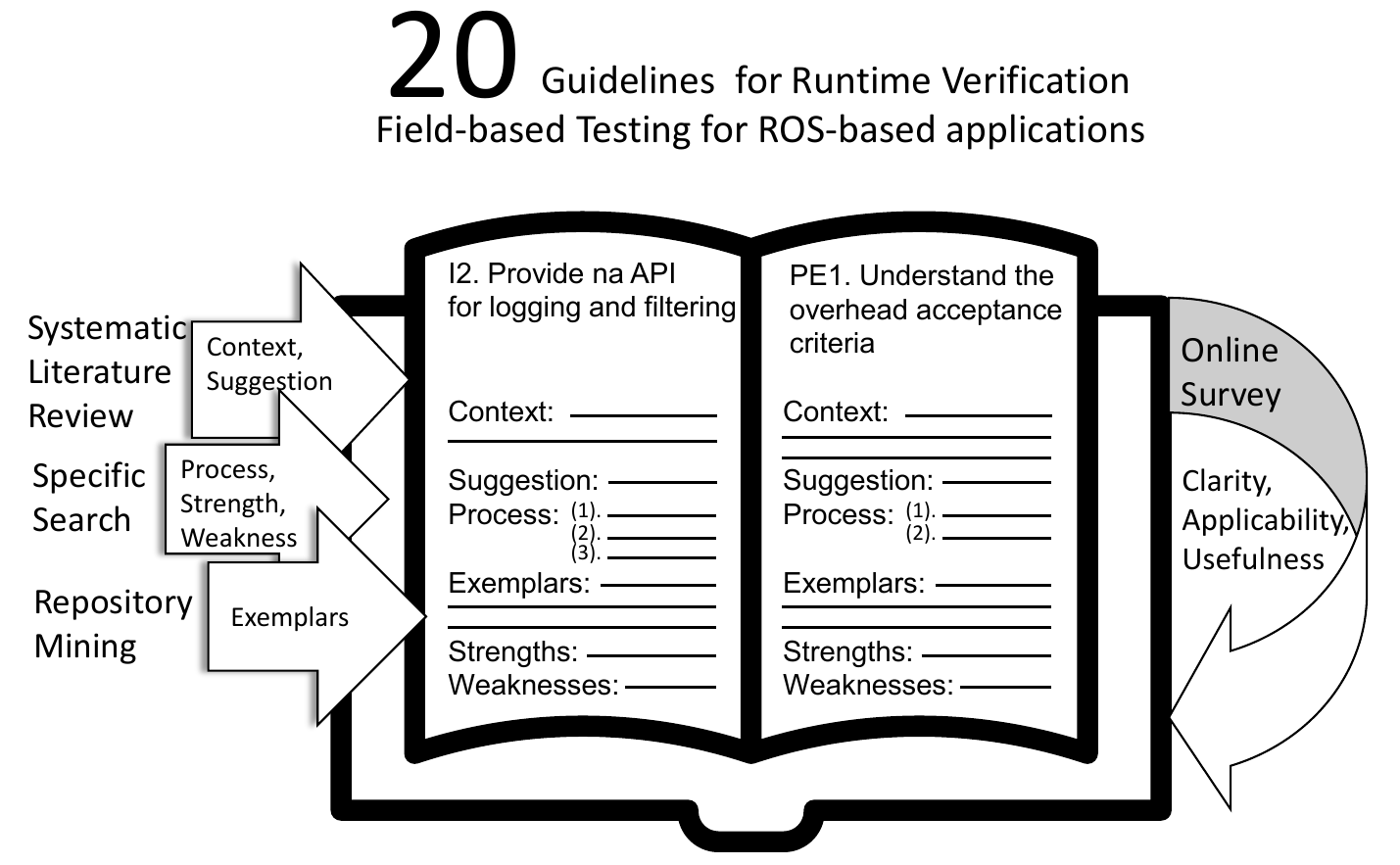}
    \caption{Synthesis of 20 guidelines to \rvft}
    \label{fig:main}
\end{figure}
 
Figure~\ref{fig:main} illustrates our methodology. First, we systematically reviewed the literature on runtime verification and field-based testing of ROS-based systems. After we sketched the first drafts of the guidelines, we performed specific searches on search engines and scientific databases tailored to the specific guidelines to complement knowledge gaps. We also performed an in-depth analysis of repositories with open-source robotics applications that contain solutions to the challenges we identified. Thereafter, we synthesized actionable guidelines and enriched by concrete exemplars collected from the repositories through an online questionnaire. Finally, we validated the guidelines' applicability, clarity, and usefulness with robotics experts, including active researchers and practitioners. 

Our study results in 20 guidelines: 8 guidelines for developers and 12 guidelines for QA teams. The catalogue of guidelines is specifically designed to instruct practitioners in the field-based testing of ROS-based systems, from both rigorous development and real-world testing and verification lenses. These guidelines are carefully crafted to address the challenges and requirements of testing and verification of robotic systems in real-world scenarios. Practitioners can identify best practices and recommendations to simplify, facilitate, and make more effective the verification and validation at runtime of their ROS-based robotic systems. 
We offer an overview of existing methods and tools for testing ROS-based systems, focusing on how they address the challenges of field-based testing and runtime verification.

Researchers can identify limitations, strengths, and gaps in the research landscape. 
We relate the guidelines with open challenges in runtime verification and field-based testing, and discuss promising research directions in the field. 

\section{Background}

We introduce the Robot Operating System (ROS) (\cref{sec:ROS}), runtime verification (\cref{sec:RVrobotics}), and field-based testing in robotics (\cref{sec:FBTrobotics}). 

\subsection{Robotic Systems Development Based on ROS}\label{sec:ROS}

Robot Operating System (ROS) is an ecosystem for developing robotic systems. Mainly, ROS comprises a middleware for interfacing fundamental robotic components (i.e., hardware, or software) with the ambition to eliminate re-inventing the wheel as a practice in robotics software engineering. ROS is used both in academia for teaching and research, and in industry, promoting commercial-friendly policies. ROS is prominently backed by a crescent and active open-source community. In the ROS ecosystem, the software is organized into basic building blocks called packages, that are meant to serve enough functionality to be useful. Over time, distributions became very large, surpassing 4000 packages in the ROS distribution Melodic Morenia. Furthermore, many open-source packages are not part of the official distributions.

A typical ROS-based system contains distributed resources interacting with each other. This network of resources is called the {\it ROS Computation Graph}, which is composed by four main elements:

\begin{itemize}
    \item {\bf Nodes} are the main resources in a ROS computation graph. They are processes that consume, process, and produce data. They communicate with each other via message passing. The other types of resources either hold shared data (parameters) or serve as message-passing channels (topics and services). Nodes should be specific and modular, rather than monolithic components. It is normal for a single robot to have a network of many nodes.
    \item {\bf Topics} is the most common message-passing mechanism. It follows an asynchronous publish-subscribe model with many-to-many connections. Publishers can send messages at any time, independently of the number of active subscribers. Subscribers, upon receiving a new message, are alerted through a callback function. Both publishers and subscribers utilize a message queue, the size of which is user-determined.
    \item {\bf Services}, which are offered as the second message-passing option by ROS, facilitate synchronous one-to-one interactions through remote procedure calls. This model distinguishes between a server (node providing the service) and a client (node using the service).
    \item {\bf Parameters}  represent the last resource type, providing data sharing among nodes without relying on direct messaging or explicit communication. ROS maintains a local parameter server, a shared key-value store, which is accessible for reading and writing by any node or user, such as through a command line.
\end{itemize}

ROS is currently migrating to ROS 2, which introduces significant changes to managing the computation graph. This requires effort from developers to migrate their applications. As a result, some providers maintain their applications in previous ROS distributions while implementing new ones in ROS 2. For instance, MoveIt, a popular framework for robotics manipulation, maintains two versions. In our paper, we distinguish between ROS and ROS 2 when necessary. We believe that eventually ROS~2 will become the standard.

\subsection{Runtime Verification in Robotics}\label{sec:RVrobotics}

Runtime Verification (or Runtime Monitoring\footnote{There is no agreement in the community about which terminology is correct.}) comprises methods to analyze and check the dynamic behavior of computational systems~\cite{havelund2005verify,bartocci2018introduction}. Runtime verification ensures that a property expressed with a formal language is not violated at runtime. It consists of a lightweight, yet rigorous, formal method that complements classical exhaustive verification techniques (e.g., model checking and theorem proving) with a more practical approach that analyses execution traces. At the price of limited execution coverage, RV can give precise information on the system's runtime behavior. 

We follow the taxonomy of Falcone et al.~\cite{falcone:2021} to discuss runtime verification in robotics. 
Runtime verification plays an important role in checking robotic software. Typically, robotic applications operate in uncertain environments, placing fundamental barriers to exhaustive verification. As such, runtime verification turns out to be a promising direction for gathering confidence in robotic systems.

In robotics, runtime verification may rely on reactive synthesis for generating monitors from specified system  properties~\cite{azzopardi2022runtime}. Monitors check properties at runtime. For example, monitors can be used to encode safety constraints that, upon violation, may trigger reactive behaviors that engage the system in a recovery mode~\cite{rizwan2023rossmarie}. Monitors check properties against traces of execution, i.e.,~a finite sequence of observations that represents the behavior of interest. The process of adding monitors to the system is called instrumentation. For instance, instrumentation may be needed to collect UAV data (e.g., GPS coordinates, attitude, and mission status), which is aggregated, analyzed, and persisted~\cite{vierhauser2023grum}. The aggregated data can either be checked on-the-fly (i.e., online monitoring) or in a post-mortem analysis (i.e., offline monitoring). The post-mortem analysis is an after-the-fact analysis technique that uses collected traces to validate the robotic system, e.g., Brunner et al.~\cite{brunner2018design} proposes an architecture to robotics with an emphasis on post-mortem analysis by combining execution history logs (a.k.a. traces), Gantt charts, resource usage profiles, and task execution metrics and statistics.


\subsection{Field-Based Testing in Robotics}\label{sec:FBTrobotics}


Field-based testing\footnote{Please notice that Field Testing is different from Field-based Testing. The former is part of the latter.} is a testing technique that uses (but is not constrained to) information from the field (a.k.a. the real world). We follow the definitions from Bertolino et al.~\cite{bertolino:2021}. As shown in Fig.~\ref{fig:bert}, field-based testing includes \emph{in-vivo} testing, i.e., tests that are executed in the production environment, and \emph{in-vitro} testing, i.e., tests that are executed in the development environment but that are using data from the field. 

\begin{figure}[htb!]
    \centering
    \includegraphics[width=.7\columnwidth]{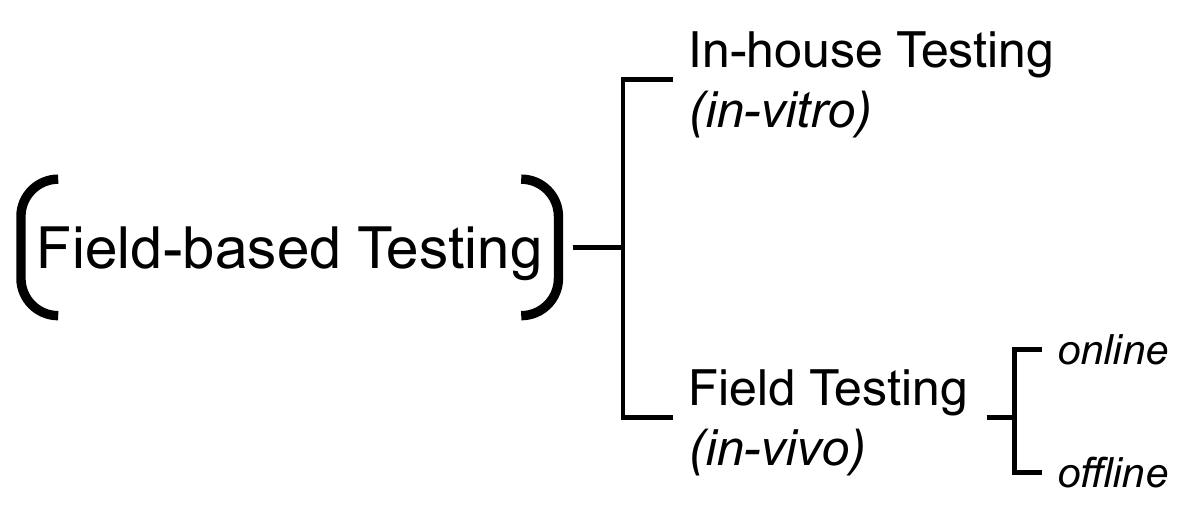}
    \caption{A classification of Field-based testing strategies by Bertolino et al.~\cite{bertolino:2021}}
    \label{fig:bert}
\end{figure}

\noindent\textbf{In-house Testing (in-vitro)} uses information from the field in tests conducted in the development environment. A typical scenario of in-vitro testing is when developers collect an abnormal behavior from runtime and want to better understand it by experimenting with such a scenario in simulation. 

\vspace{1mm}
\noindent\textbf{Field Testing (in-vivo)} refers to tests that are executed in the production environment. Field tests may be divided into two sub-categories: \textit{offline testing} and \textit{online testing}.  Field testing, as opposed to in-house testing, is desirable when the system itself operates in environments subject to uncertainty, whereas modeling and simulating the production environment is error-prone and may lead to false results. 

\vspace{1mm}
\noindent\textit{Offline Field Testing} is performed in the production environment on an instance of the system under test (SUT) that is different from the operational one.
Then, the testing team may understand and validate the system by applying the desired inputs in the replica. This approach minimizes interference in the nominal operation of the system and may be more cost-effective than online field testing.

\noindent\textit{Online Field Testing} is performed in the production environment. Online field testing is closer to reality, reducing the effect of uncertainty in the tests substantially. Other implications, such as security, privacy, and safety may emerge from performing online field testing.
\section{Methodology}\label{lb:methodology}

Our study contributes guidelines for development and quality assurance involved with engineering ROS-based applications. 
We focus on runtime verification and field-based testing to validate ROS-based applications. Our target audience is mainly practitioners and researchers interested in understanding the state-of-the-art of testing and RV in ROS in the light of growing recognition of software pivotal role in robotic software engineering. We aim to deliver concrete scientific artifacts that can be used by the ROS community~\cite{alami2018}. 

We follow \emph{design science}~\cite{wieringa:2009, baskerville:2018, hevner:2021} to answer our research questions. We perform three research cycles, as depicted in Fig.~\ref{fig:methodology}. 
Each research cycle focuses on creating awareness about the problem domain and possible solutions, synthesizing a solution, and validating whether the solution mitigates the problem. In the three cycles, we incrementally synthesized a set of guidelines that answer our research questions.

\begin{figure}[htb!]
    \centering
    \includegraphics[width=\columnwidth]{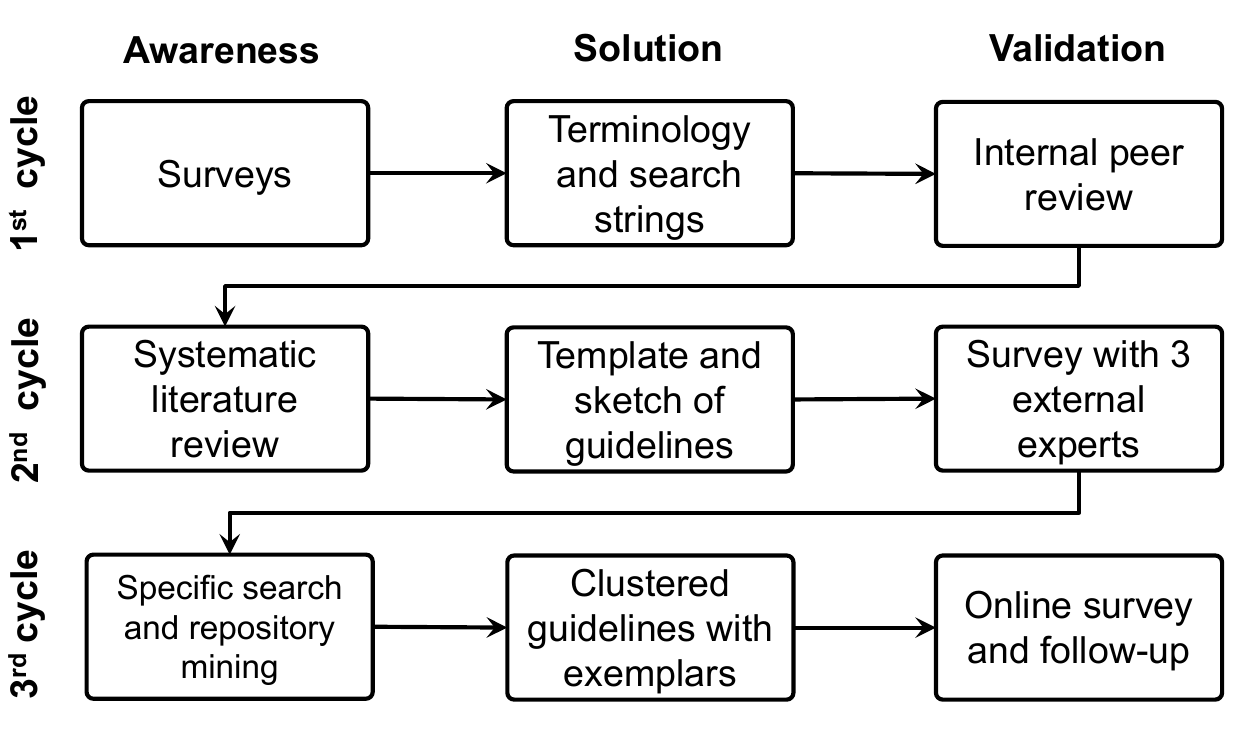}
    \caption{\label{fig:methodology} Overview of the activities to synthesizing guidelines according to design science}
    
\end{figure}

\textbf{First Cycle.} The first cycle is devoted to establishing awareness of the state-of-the-art of field-based testing and runtime verification in robotics through existing surveys. Through a literature review in Scopus, IEEE Xplore, and ACM DL, we collected surveys that discuss runtime verification and field-based testing. 
We, then, reviewed the surveys to define the terminology to delimit the scope of the study and the search strings that we would use in the next iteration. These surveys also played a role in the identification of sketches of guidelines (which should be considered as input for the second cycle) and in the identification of challenges, which are discussed in~\cref{sec:discussion}. The terminology and the search strings were internally validated through peer reviews and discussions with co-authors to consolidate the terminology and scope.

\textbf{Second Cycle.} We, then, used the search strings defined in the previous cycle to perform a systematic literature review.  Then, we defined a template for describing the guidelines and a first draft of the guidelines. More specifically, the data extracted from surveyed papers has been used to identify an initial set of guidelines. In this step, we identified 8 guidelines, 4 concerning the developers' activities and 4 concerning the quality assurance team activities (as reported in the replication package~\cite{guidelines}). As presented later in the paper, the final set of guidelines is much larger (20 guidelines) since some of this initial set of guidelines have been split into more guidelines and new ones have been added. As a preliminary validation step, we presented a subset of the guidelines' sketches at the Robotics Software Engineering workshop\footnote{\href{https://rsemeeting.github.io/rse2022/}{https://rsemeeting.github.io/rse2022/}}. We received feedback on the guidelines from three  
experts in robotics through a questionnaire and subsequent discussion.

\looseness=-1
\textbf{Third Cycle.} The third cycle was devoted to consolidating the set of guidelines and validating them with experts from industry and academia. 
The comments from the expert evaluation in cycle 2 suggested to further extend the set of guidelines due to potential missing guidelines and to complement the guidelines with concrete examples, e.g., {\it ``Can you provide code examples?''} or {\it ``Without examples I wonder how such a specification pattern could look like.''}. Thus, guided by such comments, we decided to perform a specific search and to mine online repositories. The rationale is that the SLR should be complemented by specific searches and repositories mining focused on finding additional papers, information, and concrete examples in a specific topic or argument. For example, in the initial set of 8 guidelines, we had only one guideline concerning the specification of properties. The feedback from one of the three external experts highlighted that there was the need of more precision and standardized terms. Then, thanks to specific searches and mining of repositories for property specification in ROS-based systems, the single guideline on the specification of properties has been refined into three guidelines, one focused on logic-based languages, one on domain specific languages and the third on scenario-based specification  (SDB1, SBD2, and SBD3).  
It is important to highlight that these specific searches cannot be substituted by a refined SLR including more terms (i.e., re-performing the second cycle of the methodology), since the granularity of specific searches is different from the granularity of the SLR.

To perform the specific searches and repositories mining, we used the terminology from the second cycle and the feedback from the experts. Information on the specific searches we made is available in the replication package~\cite{guidelines}; specifically, within the protocol for repositories mining, we provide tables with the keywords used for the mining activity performed for each guideline. We searched for repositories mentioned in the documents in complement to repositories that touched upon the theme of each guideline sketch. In the solution phase, we then developed the final version of the guidelines, provided exemplars for each of them, and clustered the guidelines as they are presented in this paper. Finally, we validated the final set of guidelines with practitioners and researchers. We designed and distributed an online questionnaire to experts from industry and academia, through the ROS community forum (namely ROS Discourse), LinkedIn, X (Twitter), and to experts we met at robotics and software engineering conferences. Before releasing the questionnaire to the wider community, we performed a pre-study with an experienced roboticist and active member of the ROS community and two members with widely recognized contributions to robotic software engineering from academia. The results from the third cycle confirmed the relevance of the guidelines and helped us to fine-tune them. This will be further detailed in Sect.
~\ref{sec:validation3}.


In the reminder, we detail the steps that require further explanation: the Systematic Literature Review of the second cycle is described in \cref{sec:slr}; the specific search and repository mining step is described in \cref{sec:ss-mr}, and the online survey
and follow-up step of the third cycle is described in \cref{sec:validation3}. 
All generated artifacts and information are available in the replication package~\cite{guidelines}. 

\subsection{Systematic Literature Review}\label{sec:slr}

Given the lack of comprehensive scientific reports on runtime verification and field-based testing in robotics, we used a systematic literature review to identify the available and relevant research~\cite{kitchenham2009systematic}.
To ensure a clear and thorough methodology, we established a protocol for our literature review, outlining our research goals, the process to be followed, data extraction, and measures to mitigate threats to the validity of our results.  Detailed information is provided in the replication package~\cite{guidelines}. 

\noindent\textbf{Research Goal.} Characterize runtime verification (RV) and field-based testing (FT) for building confidence in ROS-based applications from the researchers' perspective.

\noindent\textbf{Process.} Our search process started by defining relevant research questions, which informed the creation of a search string to be used in automatic search engines (i.e., IEEE Xplore, ACM Digital Library, and Scopus). The search string was composed of terms elicited from previous studies on runtime verification tools~\cite{falcone:2021} (lines 3-4) and field-based testing~\cite{bertolino:2021} (lines 6-8).

\lstdefinelanguage{SLR}
             { 
             identifierstyle=\color{black},
            keywordstyle=\bfseries,
            keywords={},
            otherkeywords={),*,(,OR,AND}
            }

\lstset{frame=tb,
language=SLR,
                    breaklines=true,
                    columns=flexible,
                    numbers=none,
                    xleftmargin=0cm,frame=tlbr,framesep=4pt,framerule=0pt
                    }

\begin{lstlisting}[basicstyle=\fontsize{\f@size}{\f@baselineskip}\footnotesize]
(ros OR robot* operating system) AND
(runtime verification OR run-time verification OR
 runtime assurance OR run-time assurance OR 
 online verification on-line verification OR 
 runtime monitoring OR run-time monitoring OR
 runtime testing OR run-time testing OR 
 online testing OR  on-line testing OR 
 field-based testing OR field testing OR in-vivo testing)
\end{lstlisting}

As shown in Figure ~\ref{fig:lit_rev}, the search resulted in 262 papers. We applied pre-filters in the search engine to only include papers from Computer Science and Engineering and similar areas, reducing the number of studies to 132 merged and unique papers. Then, we used inclusion and exclusion criteria for screening for relevance, resulting in 27 relevant studies. 

\begin{figure}[tb!]
    \centering
    \includegraphics[width=\columnwidth]{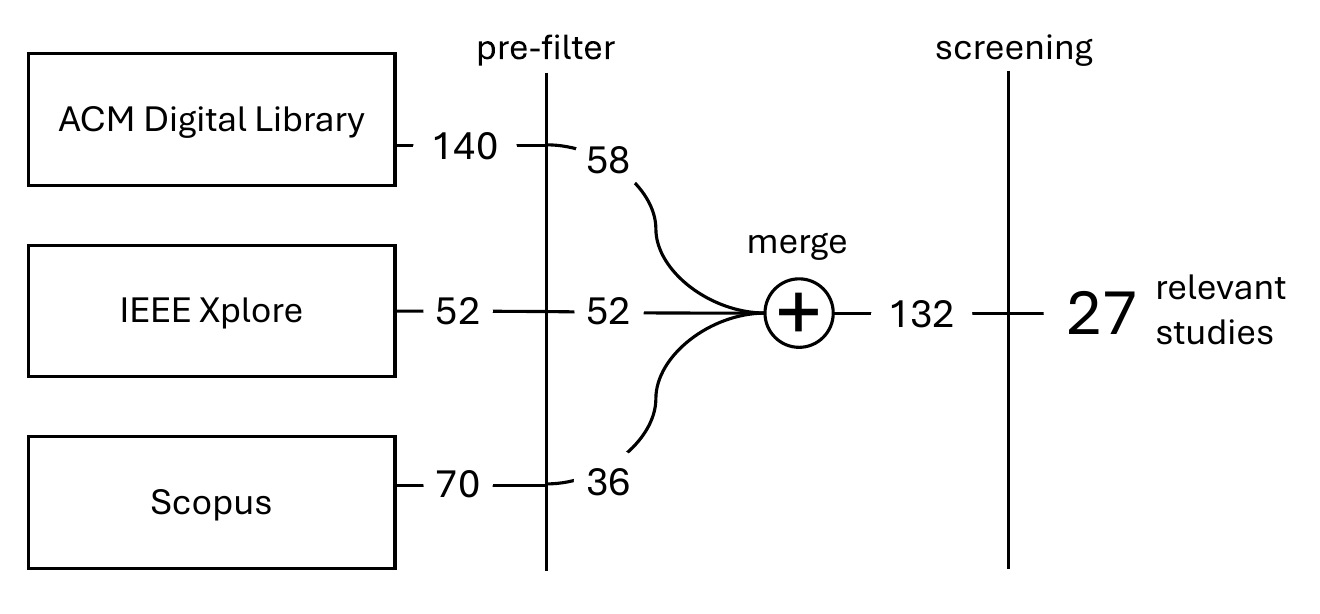}
    \caption{\label{fig:lit_rev} Literature review and quantities of studies}
\end{figure}

The inclusion criterion tailors the study to ROS, independently of the version or distribution, and the exclusion criteria determine that studies should not be short papers, secondary studies, or duplicated. We also exclude papers that are not from Computer Science or Engineering.

\noindent\textbf{Data Extraction.}
The screening phase and the research goals were the basis for defining a classification scheme as a set of categories that should be used for extracting data from the resulting studies. The categories used in the scheme define the meta-information of the publications, the nature of the assurance technique employed (i.e. runtime verification or field-based testing), a classification of the applied techniques according to the taxonomies for runtime verification~\cite{falcone:2021} and field-based testing~\cite{bertolino:2021}, the software quality attributes assessed in the studies, use cases, motivation, and rationale. Finally, if the study needed to be more precise about how they used RV or FT, we would query about the used technique, whether it is static or dynamic, and whether it was performed offline or online. With the classification scheme and the studies in hand, data extraction is a matter of carefully reading and collecting information from the studies. The extraction activity resulted in 9 studies reporting on field-based testing, 16 studies reporting on runtime verification, 1 study discussing both, and 2 studies discussing none of them, but, instead, providing valuable insights into runtime verification and field-based testing in ROS.

\subsection{Specific search and repository mining}\label{sec:ss-mr}

Guided by the comments from experts (2nd cycle), we (three authors of this paper) searched for additional information to complement the guidelines. As defined in the description of the third cycle, we focused on finding additional papers and concrete code examples to make guidelines more useful.
We performed both a specific search\footnote{The search was performed in December 2023} and mined online repositories. First, for the specific search, we manually selected results searched from Google Scholar and the ACM digital library. 
Then, we followed the mining approach by Malavolta~\cite{malavolta2021mining} to collect new online repositories. We collected and cloned 1088 repositories, searched for patterns in the files within the repositories, performed a manual context analysis. The results went through a peer review with the other authors of our paper. The peer review aimed at a refinement of the search, or at potentially finding new exemplars. Further information and the scripts described in this section are available in our online replication package~\cite{guidelines}.

To collect and clone the 1088 repositories we relied on rosmap,\footnote{\href{https://github.com/jr-robotics/rosmap}{https://github.com/jr-robotics/rosmap}} a tool designed for dependency mapping, which allowed the gathering of an exhaustive list of all public ROS projects available on Github, as well as Bitbucket and Gitlab. Then, to filter out unwanted repositories and download respositories to a local server, we modified the Python script from Malavolta's replication package.\footnote{\href{https://github.com/S2-group/jss-2021-replication-package}{https://github.com/S2-group/jss-2021-replication-package}}

\begin{itemize}
    \item {\it explorer.py} removes repositories that are duplicate, forked, lower than 100 commits, have a low star count ($\leq 2$), and are demo projects; such repositories have a low relevance to our study, which looks into exemplars to complement the guidelines.
    \item {\it cloner.py} downloads the repositories to a local server. The local setup allows for quick targeted searches inside the source files of mature projects, with a high degree of search-term flexibility.
\end{itemize}

We curated the downloaded repositories by manual inspection. We used the UNIX 'grep' utility to efficiently scan large numbers of files.
Particularly for guidelines with no straightforward exemplars, we searched for exemplars by looking up keywords relevant to each of the guidelines and examining code comments in the files (e.g., LLVM, LibFuzzer, which we found while looking for fuzzing mechanisms, used by PX4 Autopilot, a large open source project~\cite{PX4}).
Moreover, for guidelines that already contained exemplars from the literature, we performed dependency analysis by looking into source files of the other repositories, specifically in their {\it import} statements.
Whenever we found matching files, either by keyword search or dependency analysis, we analyzed peripheral code which helped us to discard irrelevant exemplars. Finally, the remaining repositories went through peer-review sessions in which other authors of this paper determined whether the repositories were viable exemplars discovered for the corresponding guidelines.

\subsection{Online Survey and Follow-up}\label{sec:validation3}

Although our guidelines are grounded on data, provenient from the scientific literature and mined repositories, we made an additional validation step with experts in robotics particularly in runtime verification, or field-based testing to validate the adherence of the guidelines with the target groups. The validation aims at collecting the practitioner's sentiment on whether each guideline is useful, clear, and applicable. The instrument of this validation was online questionnaires with follow-up emails. Our validation strategy comprised (i) recruitment, (ii) questionnaire design, (iii) results analysis, and (iv) follow-ups and reporting data.

\subsubsection{Recruitment: Collecting Names and Contacts of Robotics Experts}\label{sec:recruit}

The guidelines intend to contribute practical suggestions on how (i) developers should design robotic systems to enable runtime verification and field-based testing, and how (ii) quality assurance teams can verify and test ROS-based applications. 
We targeted practitioners, i.e., developers and QA teams with experience in robotics (ROS specifically), runtime verification~\cite{falcone:2021}, field-based testing~\cite{bertolino:2021}, and researchers with experience in robotics.
We used three data sources to collect names and contacts: authorship and email addresses in the papers and repositories used as sources of information for the synthesis of the guidelines, advertisement in social media and forums, and renowned experts in software robotic research.

Overall, our survey received 55 answers to the questionnaire. We manually collected names and emails from papers, which resulted in 306 contacts. Our mining algorithm provided 772 emails from the authors from mined repositories. In addition, we shared the questionnaire in the ROS discourse forum.\footnote{\href{https://discourse.ros.org/t/we-need-participants-for-a-survey-on-field-based-testing-of-ros-applications/34879}{https://discourse.ros.org/t/we-need-participants-for-a-survey-on-field-based-testing-of-ros-applications/34879}} By March of 2024, the post was accessed by 1.6k ROS discourse users. Moreover, the link to the questionnaire targeting developers was accessed 60 times and the link to the questionnaire targeting quality assurance teams was accessed 8 times. OpenRobotics also shared the discourse post on LinkedIn\footnote{\href{https://www.linkedin.com/feed/update/urn:li:activity:7138595420783484929/}{https://www.linkedin.com/feed/update/urn:li:activity:713859\\5420783484929/}} and X (Twitter).\footnote{\href{https://twitter.com/OpenRoboticsOrg/status/1732832183507951929?t=_bBFf7RptL7QM3wrEQ0KnA&s=19}{{https://twitter.com/OpenRoboticsOrg/status/17328321835079\\51929?t=\_bBFf7RptL7QM3wrEQ0KnA\&s=19}}}

We also participated in conferences, meetings, workshops, and summer schools during the study. We invited a list of 67 robotics experts to participate in this survey. To avoid bias, before inviting candidate respondents, we did not discuss the subject matter of any of the guidelines but discussed the idea of synthesizing guidelines to support \rvft~of ROS-based applications.

Our recruitment campaign started on November 29th 2023 and ended on the 5th of January of 2024, when we were not receiving any more answers from the candidate respondents.

\subsubsection{Questionnaire Design}

The survey aims to capture the target group's opinion and experience in runtime verification and field-based testing ROS-based systems.
Since we decided to separate the guidelines targeting to two different audiences, developers and QA teams, we also decided to use two separate questionnaires. First, the introduction sets the context and gives instructions on how to answer the questions. Then, the main part for each guideline consisted of three questions eliciting usefulness, clarity, and applicability. We also included open-ended questions that enabled respondents to express their opinions about one or more guidelines. Third, the profiling section asked about the respondents' background and previous experience with software engineering for robotics, especially ROS-based applications.

We piloted the questionnaire design with a member of the ROS community, who was responsible for maintaining the ROS Discourse Forum,\footnote{\href{https://discourse.ros.org/}{https://discourse.ros.org/}} and with an additional expert from academia. In addition, we contacted a professor with over ten years of research on robotics software engineering, and a postdoc that has recognizable contributions to the robotics software engineering field. The main comments from the pre-study highlighted that (i) the term ``field-based testing'' was not clear, (ii) some guidelines were not guidelines, and (iii) the guidelines were written too formally, with {\it academic jargon}. We addressed their comments and rectified the questionnaire with the same member who assisted in publishing the post.


\subsubsection{Result Analysis}

We mainly relied on quantitative data analysis combined with statistical tests to help us build confidence in our results and steer follow-up analyses. 
To analyze the Likert questions we used (i) Likert plot visualizations of responses and (ii) quantitative statistical analyses, i.e., Boxplot, significance test. We used the analysis to follow-up with the respondents to better understand variance in the responses on specific guidelines. To perform the statistical analyses (ii), we assume that the Likert items can be interpreted as intervals, and we use consistent confidence levels ($\alpha=0.05$) and the Cohen's~r
parameters to analyse the effect size. 
\section{Guidelines}

\begin{figure*}[htb!]
    \centering
    \includegraphics[width=\textwidth]{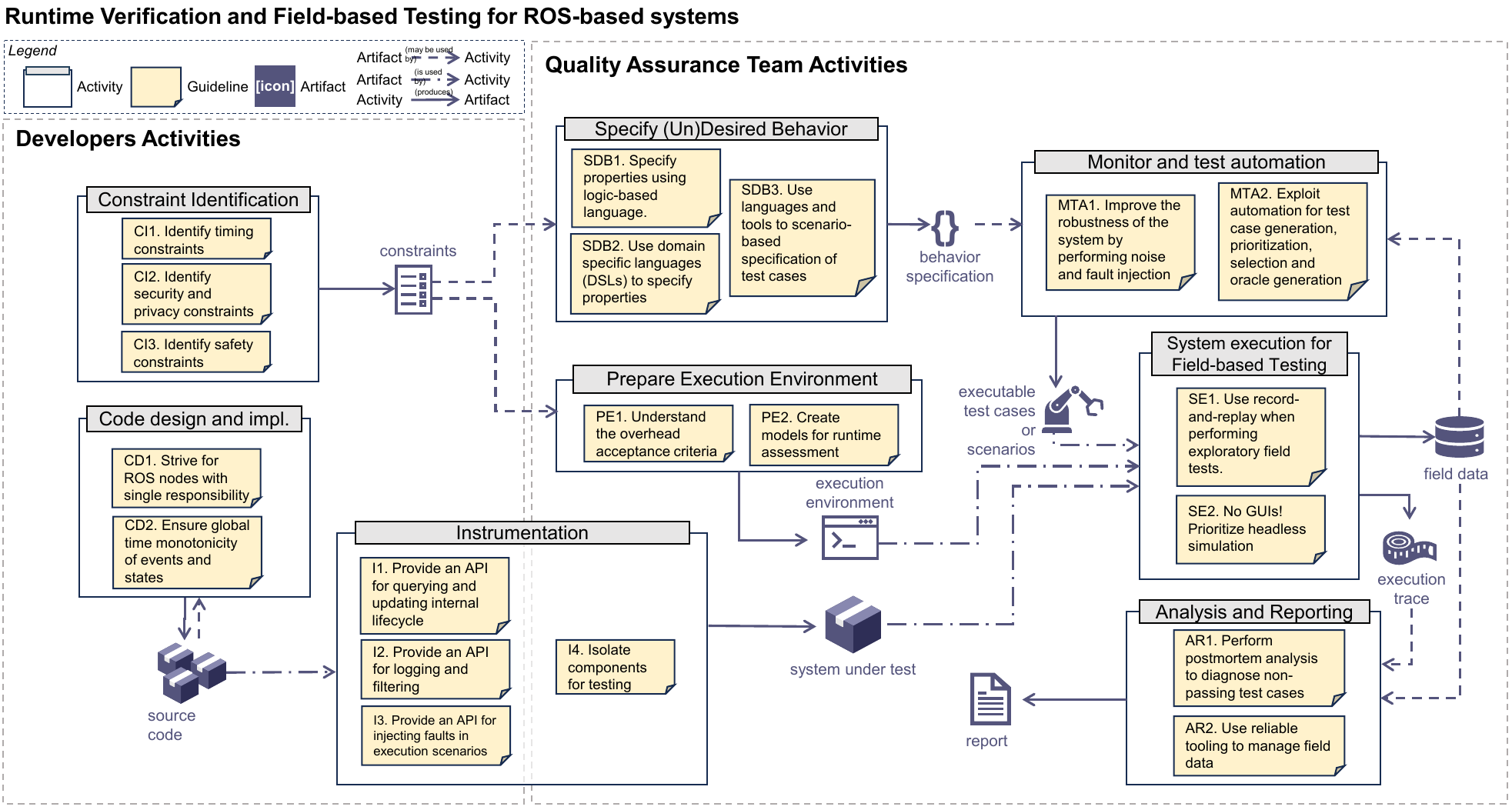}
    \caption{Overview of the guidelines organized according to the engineering activities of developers and the QA teams when engineering ROS-based software and who strive to establish field-based testing and runtime verification}
    \label{fig:dev-test-process}
\end{figure*}

We now describe the guidelines we synthesized in our study. \cref{sec:overview} provides an overview of the guidelines for preparing and performing \rvft{}~for ROS-based systems. Then, before describing the guidelines in detail, \cref{sec:template} describes the template used to specify guidelines. Finally, \cref{sec:guide_facilitate} and~\ref{sec:guide_testing} highlight two representative guidelines. The complete list of guidelines can be found on our dedicated website~\cite{guidelines}.

\subsection{Guidelines Overview}\label{sec:overview}
Figure \ref{fig:dev-test-process} provides an overview of the guidelines. 
The guidelines are organized into two groups: (i) guidelines concerning activities that are performed to prepare ROS-based systems for field-based testing and to enable runtime verification, and (ii) guidelines concerning the actual \rvft{}~of ROS-based systems. The first set of guidelines is mostly defined for developers, and the second set is for quality assurance (QA) teams. Then, guidelines are organized into eight activities, each focused on specific activities: 

\begin{itemize}
    \item {\it Constraint Identification (CI)}  guidelines concerns the elicitation of  
non-functional requirements that may be critical or non-negotiable to the system under design and test. Preparing or running test scenarios to attest correct behavior of the system should not violate the identified constraints. These constraints might be useful for the {\it Specify (Un)Desired behavior} SDB group of guidelines (explained shortly); however, the specification of properties or test cases can use other sources of information, e.g., a requirement document. 
\item \textit{Code Design and Implementation (CD)} includes guidelines 
that may facilitate testing and verification by modifying the source code of the robotics software under scrutiny~\cite{luckcuck:2019,proetzsch2007behaviour}. 
\item \textit{Instrumentation for \rvft (I)} includes guidelines to support either the developers or the testers in instrumenting the code resulting in a system ready to test. Instrumentation in this context refers to modifying the source code to expose internal variables to monitors or oracles~\cite{falcone:2021,falcone2013tutorial,bartocci2018introduction}. 
\item \textit{Specify (Un)Desired behavior (SDB)} 
takes into consideration the elicited constraints and provides the means to abstract the system behavior in terms of states and events (a.k.a. state changes) and how specification languages can be used to describe properties of a set of such states or events~\cite{falcone:2021,bartocci2018introduction}.
\item \textit{Monitor and Test Automation (MTA)} concerns guidelines to automate the generation of monitors or test, including the synthesis of test cases or test scenarios that may accept or reject an observed execution trace~\cite{falcone:2021,falcone2013tutorial,bartocci2018introduction,bertolino:2021}. 
\item \textit{Prepare Execution Environment (PE)} concerns guidelines to prepare the environment by setting up supporting devices such as stubs and mocks to \rvft~\cite{luckcuck:2019,proetzsch2007behaviour,falcone2011runtime}. The environment preparation may take into consideration testing constraints previously elicited to avoid damages to the test or lead to skewed results. 
\item 
\textit{System Execution for Field-based Testing (SE)} includes guidelines concerning the activity of running one or more test cases or test scenarios in a given execution environment to exercise a system under test, which results in field data~\cite{falcone:2021,bertolino:2021,falcone2013tutorial}. 
\item \textit{Analysis and Reporting (AR)} gathers the field data generated during test execution and derives evidence, conclusions, and finally a report on the observed behavior~\cite{falcone:2021,bertolino:2021,falcone2013tutorial}. In the case of inconclusive results, the report may be used by the developers to further refine the constraints or code design. Otherwise, the report may be used as a collection of assurance arguments to leverage confidence in the tested ROS-based system.
\end{itemize}

 
It is important to highlight that the overview offered in Fig.~\ref{fig:dev-test-process} does not define a new development process for ROS-based systems. The guidelines are agnostic to the development process used. Moreover, the guidelines are independent of each other, and each of them can be used in isolation. Also, it is the responsibility of the developers or the QA team to select and use those guidelines that are more beneficial for the system they are developing and/or analysing. However, as shown Fig.~\ref{fig:dev-test-process} groups of guidelines are connected by arrows, since the artifact produced by some guidelines might be useful for other guidelines. Specifically, the relationships among guidelines are of three different types: (i) an artifact {\it might be used} by an activity, (ii) an artifact {\it is used by} an activity, and (iii) an activity {\it produces} an artifact. In the legend of the figure, we explain the graphical representation used for these three types of relationships.

The guidelines are detailed in \cref{sec:guide_facilitate} and~\ref{sec:guide_testing} after the description of the template used to document them (in \cref{sec:template}). Our dedicated website~\cite{guidelines} contains the full list of guidelines for enabling and facilitating runtime verification and field-based testing of ROS-based systems.


\subsection{Template for Guideline Specification}\label{sec:template}

In the following, we describe the fields that compose the template. The template is used in \cref{sec:guide_facilitate} and~\ref{sec:guide_testing}, as well as, to document all guidelines in our dedicated website~\cite{guidelines}.

\vspace{.5cm}
\noindent\textbf{ID:} Identifier used to facilitate tracing guidelines between groups.
    
\noindent\textbf{Title:} Title summarizes an action that practitioners should follow to mitigate or avoid a recurring problem.

\noindent\textbf{Context (WHEN):} The Context is a paragraph placing the guideline among a known set of conditions. This paragraph should delimit the scope in which the guideline is applicable. It should also introduce the conceptual terminology used in the guideline, which is defined by the conditions under which the guideline is valid. 
   
\noindent\textbf{Reason (WHY):}  Reason introduces the recurring problem faced by practitioners. It intends to leverage the relevance of the guideline to practitioners. 

\noindent\textbf{Suggestion (WHAT):}  Suggestion is a sentence or two introducing WHAT should the practitioners do to mitigate the recurring problem. 
        
\noindent\textbf{Process (HOW):}   Process is a paragraph that carefully guides the practitioners through HOW they can practice the guideline. In this paragraph, there should be references to tools that may help, concrete examples, or references to precise explanations of researchers or practitioners who have done something similar. 
    
\noindent\textbf{Exemplars:} Exemplars are concrete descriptions of papers/artifacts that follow the guidelines. An exemplar can be generic such as an artifact or rather specific such as a model problem in testing ROS-based systems in the field. 

\noindent\textbf{Strengths:} Strengths is a list of benefits that the practitioners should consider when applying the guideline. 

\noindent\textbf{Weaknesses:} Weaknesses is a list of technological/theoretical barriers that may slow the actual implementation of the guideline, either by undesired side-effects or scenarios in which applying the guideline might not lead to the desired effect.

\subsection{Guidelines for Development to Support Runtime Verification and Field-Based Testing}\label{sec:guide_facilitate}
We synthesized the guidelines in Tab.~\ref{tab:guid_sources1} relevant for enabling and facilitating runtime verification and field-based testing through \textit{development} activities, except I4 that concerns QA activities.
 The guidelines are organized in groups, as explained in \cref{sec:overview} and summarized in Fig.~\ref{fig:dev-test-process}. For each guideline we summarize the name of the guideline, and according to the main activities described in the research methodology (\cref{lb:methodology}), we provide the links to (i) the relevant initial surveys (column Initial), (ii) the papers identified during the systematic literature review (column Lit. Review), (iii) papers identified during the specific search phase (column Specific Search), and (iv) exemplars identified with the repositories mining (column exemplars). The group of Constraint Identification (CI) guidelines contains three guidelines, CI1, CI2, and CI3, devoted to identifying constraints of different nature, i.e., timing constraints, security and privacy constraints, and safety constraints. The group of Code Design and Implementation (CD) guidelines contains two guidelines, CD1 recommending to design ROS nodes with single responsibility and CD2 guiding towards ensuring global time monotonicity of events and states. The last group contains four Instrumentation (I) guidelines. I1, I2, and I3 focus on providing APIs for querying and updating an internal lifecycle (I1), for logging and filtering (I2), and for injecting faults in execution scenarios (I3). The last one (I4) focuses on isolating components for testing.
We elaborate on the guideline I2 (provision of an API for logging and filtering) in depth because we believe I2 can help understand each field of the template as well as the group of guidelines to facilitating field-based testing and runtime verification since it contains several exemplars.

\begin{table*}[htb!]
    \centering
    \caption{Guidelines to facilitating field-based testing and runtime verification of ROS-based systems}
    \begin{tabular}{llcccl}
    \toprule
    & \textbf{Guideline} & \textbf{Initial} & \textbf{Lit.Review} & \textbf{Specific Search} & \textbf{Exemplars} \\ 
    
    \midrule
    \multirow{1}{*}{\rotatebox{90}{Constraint}} \multirow{1}{*}{\rotatebox{90}{Identification}} & 
    


     \begin{tabular}{@{}l@{}} CI1. Identify timing constraints \end{tabular} & \cite{bertolino:2021,schutz2007testability} & \cite{beul:2017,lewis:2022} & \begin{tabular}{@{}l@{}} \cite{li2022autoware_perf,albonico2023software} \end{tabular} & \begin{tabular}{@{}l@{}}Autoware\_Perf\cite{li2022autoware_perf}\href{https://github.com/azu-lab/ROS2-E2E-Evaluation}{\faicon{github}}\end{tabular} \vspace{0.1cm} \\\cline{2-6} \vspace{0.1cm}
     

     &  \begin{tabular}{@{}l@{}}  CI2. Identify security and privacy\\constraints \end{tabular} & \begin{tabular}{@{}l@{}}\cite{malavolta2021mining, falcone:2021} \\ \cite{bertolino:2021, luckcuck:2019} \end{tabular} & \cite{hu:2019,huang:2014} & \cite{mayoral2022sros2,jeong2017study}  & \begin{tabular}{@{}l@{}}ROSRV\cite{huang:2014}~\href{https://github.com/cansuerdogan/ROSRV/blob/master/docs/Usage.md}{\faicon{github}}\\SROS\cite{mayoral2022sros2}\href{https://github.com/ros2/sros2}{\faicon{github}} \end{tabular} \vspace{0.1cm} \\\cline{2-6}

     & \begin{tabular}{@{}l@{}} CI3. Identify safety constraints \end{tabular} & \cite{afzal:2020,bertolino:2021}  & \cite{huang:2014}  & \cite{adam2014towards,stadler2023romosu,bozhinoski2019safety}  &  \begin{tabular}{@{}l@{}}ROMoSu\href{https://github.com/MStadler-Organization/ROMoSu}{\cite{stadler2023romosu}\faicon{github}} \\
    Kobuki\href{https://github.com/yujinrobot/kobuki}{\faicon{github}}  \end{tabular} \\

    \midrule
    \midrule
    
    \multirow{1}{*}{\rotatebox{90}{Code }} \multirow{1}{*}{\rotatebox{90}{design }}
    \multirow{1}{*}{\rotatebox{90}{and impl.}}& 
 
        \begin{tabular}{@{}l@{}} CD1. Strive for ROS nodes with single\\responsibility \end{tabular} & \cite{malavolta2021mining,bertolino:2021} & \cite{hartsell:2019,hutchison2018} & \begin{tabular}{@{}l@{}} \cite{limsoonthrakul2009modular,efatmaneshnik2017study, rovida2017skiros} \\ \cite{albore2023skill,Gil2021}  \end{tabular} & \begin{tabular}{@{}l@{}} bsn\cite{Gil2021}\href{https://github.com/lesunb/bsn}{\faicon{github}}\\skiros2\cite{rovida2017skiros}\href{https://github.com/RVMI/skiros2}{\faicon{github}} \\ hmrs\_mission\_control\cite{rodrigues2022architecture}\href{https://github.com/lesunb/hmrs_mission_control}{\faicon{github}}\end{tabular} \\ \cline{2-6}

    & \begin{tabular}{@{}l@{}} CD2. Ensure global time monotonicity\\ of events and states \end{tabular} & \cite{malavolta2021mining} & \cite{halder2017formal} & \cite{chaaban2023new,choi2021picas,blass2021ros} & \begin{tabular}{@{}l@{}} ros2-picas~\cite{choi2021picas}\href{https://github.com/rtenlab/ros2-picas}{\faicon{github}}\\mavros~\href{https://github.com/mavlink/mavros}{\faicon{github}}\end{tabular} \\ 

    \midrule
    \midrule

    \multirow{10}{*}{\rotatebox{90}{Instrumentation}} & 
    
     \begin{tabular}{@{}l@{}} I1. Provide an API for querying and\\updating  internal lifecycle \end{tabular} & \cite{malavolta2021mining,falcone:2021, bertolino:2021} & \cite{conte:2018,hutchison2018} & \cite{belsare2023micro,nordmann2021system, staschulat2020rclc}  & \begin{tabular}{@{}l@{}} Micro-ROS\cite{belsare2023micro}\href{https://github.com/micro-ROS}{\faicon{github}} \\ system$\_$modes\cite{nordmann2021system}\href{https://github.com/micro-ROS/system_modes}{\faicon{github}} \\ rclc$\_$lifecycle~\cite{staschulat2020rclc}\href{https://github.com/ros2/rclc/tree/master/rclc_lifecycle}{\faicon{github}} \\ lifecycle$\_$service$\_$client \cite{nordmann2021system}\href{https://github.com/ros2/demos/blob/foxy/lifecycle/src/lifecycle_service_client.cpp}{\faicon{github}} \end{tabular}  \\ \cline{2-6}

     & \begin{tabular}{@{}l@{}} I2. Provide an API for logging and\\filtering \end{tabular} & \cite{afzal:2020,falcone:2021,bertolino:2021} & \cite{ferrando:2020} & \cite{elbaum2000software,mitrevski2022deploying}  & \begin{tabular}{@{}l@{}} rosout\href{https://github.com/ros/ros_comm/blob/noetic-devel/clients/rospy/src/rospy/impl/rosout.py}{\faicon{github}} \\ cloudwatchlogs-ros2~\href{https://github.com/aws-robotics/cloudwatchlogs-ros2}{\faicon{github}} \\ ROSMonitoring\cite{ferrando:2020}~\href{https://github.com/autonomy-and-verification-uol/ROSMonitoring/blob/master/generator/online_config.yaml}{\faicon{github}} \\ fault-tolerant-ros-master\href{https://github.com/PushyamiKaveti/fault-tolerant-ros-master/blob/master/src/ros_comm/rosmaster/src/rosmaster/master_api.py}{\faicon{github}}\\ black-box\cite{mitrevski2022deploying}~\href{https://github.com/ropod-project/black-box}{\faicon{github}} \end{tabular}  \\ \cline{2-6}


     & \begin{tabular}{@{}l@{}} I3. Provide an API for injecting faults\\in execution scenarios  \end{tabular} & \cite{afzal:2020,falcone:2021,bertolino:2021} & \cite{hutchison2018} & \begin{tabular}{@{}l@{}} \cite{duraes2006emulation,barbosa2012fault,lenka2018fault} \end{tabular}   & 
     \begin{tabular}{@{}l@{}} imfit\cite{yayan2023}\href{https://github.com/inomuh/imfit}{\faicon{github}}\\
     ros1\_fuzzer\href{https://github.com/aliasrobotics/ros1_fuzzer}{\faicon{github}}\\
     ros2\_fuzzer\href{https://github.com/aliasrobotics/ros2_fuzzer}{\faicon{github}}
     \end{tabular}  \\ \cline{2-6}

     & \begin{tabular}{@{}l@{}} I4. Isolate components for testing \end{tabular} & \cite{falcone:2021,bertolino:2021} & \cite{huang:2014,ferrando:2020} & \cite{macenski2023impact,BEDARD2023104361} &\begin{tabular}{@{}l@{}}  ROSRV~\cite{huang:2014}~\href{https://github.com/cansuerdogan/ROSRV}{\faicon{github}} \\ ROSMonitoring\cite{ferrando:2020}~\href{https://github.com/autonomy-and-verification-uol/ROSMonitoring}{\faicon{github}} \end{tabular}\\ 



    \bottomrule
    \end{tabular}%
    \label{tab:guid_sources1}
\end{table*}

\subsubsection*{Guideline I2: Provide an API for Logging and Filtering} \label{sec:I2}

The guideline I2 recommends developers provide an API for inline or outline logging and filtering to gather relevant data from runtime or the field given that the quality assurance team might not have access to the system under test. Using I2 renders effortless observability of inner states and events, but warns that the misuse of I2 may turn into performance issues, adding noise to data, false positives, or open security breaches. 

\vspace{.5cm}
\noindent {\bf {\bf ID}} I2

\noindent{{\bf Title}} Provide an API for Logging and Filtering

\noindent{{\bf Context (WHEN)}} Dynamically gathering information is a fundamental step for gaining confidence in ROS-based systems. Logging (and playback) is, in fact, one of the most used techniques for testing ROS-based systems~\cite{afzal:2020}. Often named \textit{monitoring}~\cite{falcone:2021} or \textit{logging}~\cite{bertolino:2021}, the process of recording textual or numerical information about events of interest may be a valuable input to the testing team. With such data at hand, the testing team will process the data and transform it into useful information to challenge their hypotheses about how the system should work.

\noindent{{\bf Reason (WHY)}} Logging important events depends on instrumenting the code (with `hooks') that enables the monitoring or logging. It is unrealistic to assume that the testing team will have access to the source code or that the testing team knows what events to log or how to do so.

\noindent{{\bf Suggestion (WHAT)}} 
The development team should provide an API for logging and filtering data to enable access to valuable runtime data used for runtime verification and field-based testing. The standard for logging and filtering is rosbag (\href{http://wiki.ros.org/rosbag}{wiki: rosbag}). Though, in addition, AWS CloudWatch (\href{https://github.com/aws-robotics/cloudwatchlogs-ros2}{git: aws-robotics/cloudwatchlogs-ros2}) collects data from the rosout topic and provides a filter for eliminating noise from the logged events. Another example is the Robotic Black Box (\href{https://github.com/ropod-project/black-box}{git: ropod-project/black-box}) which allows for listening to data traffic from distinct sources and logging the messages using MongoDB.

\begin{table*}[htb!]
    \centering
    \caption{Guidelines for performing \rvft ~on ROS-based applications}
    \begin{tabular}{llcccl}
    \toprule
    & \textbf{Guideline} & \textbf{Initial} & \textbf{Lit.Review} & \textbf{Specific search} & \textbf{Exemplars} \\ 
    \midrule
    
    \multirow{5}{*}{\rotatebox{90}{Prepare}}
    \multirow{5}{*}{\rotatebox{90}{Execution}}
    \multirow{5}{*}{\rotatebox{90}{Environment}}& 

    \begin{tabular}{@{}l@{}} PE1. Understand the overhead \\ acceptance criteria \end{tabular} & \cite{malavolta2021mining, falcone:2021, bertolino:2021} & \cite{rivera:2021,ferrando:2020} & \cite{bertolino2011enhancing,li2022autoware_perf, lahami2016safe, breiling2017secure}  & \begin{tabular}{@{}l@{}} ROS-Immunity~\cite{rivera:2021} \\ ROSMonitoring~\cite{ferrando:2020}\href{https://github.com/autonomy-and-verification-uol/ROSMonitoring}{\faicon{github}} \\ ros2$\_$tracing\cite{BEDARD2023104361}\href{https://github.com/ros2/ros2_tracing}{\faicon{github}}~\\
    RTF4ADS~\cite{lahami2016safe} and \\ Secure Channels in ROS~\cite{breiling2017secure}\end{tabular}    \\\cline{2-6}
    
     & \begin{tabular}{@{}l@{}} PE2. Create models for runtime \\assessment \end{tabular} & \cite{bertolino:2021} & \begin{tabular}{@{}l@{}}\cite{guerrero:2021,lewis:2022,grigoropoulos2020simulation}\\ \cite{durschmid2024rosinfer,cheng2020ac, pelletier2023predictive}\end{tabular} &  \begin{tabular}{@{}l@{}}\cite{schlegel2009robotic,kirchhof2020model,sotiropoulos2017can, saavedra2022ros} \\ \cite{fend2022cpsaml, ernits2015model, larsen2005testing, garcia2019boot1} \\ \cite{hammoudeh2021} \\
      \end{tabular}& \begin{tabular}{@{}l@{}} ROS$\_$Tecnomatix~\cite{saavedra2022ros}\href{https://github.com/INTELYMEC/ROS_Tecnomatix}{\faicon{github}} \\ cpsaml\cite{fend2022cpsaml}~\href{https://github.com/me-big-tuwien-ac-at/cpsaml}{\faicon{github}}\\
        dtron + adapter~\cite{ernits2015model,larsen2005testing}\\ ros-model~\cite{garcia2019boot1,hammoudeh2021}\href{https://github.com/ipa-nhg/ros-model}{\faicon{github}}\end{tabular}  \\

    \midrule
    \midrule

     \multirow{5}{*}{\rotatebox{90}{Specify (Un)-}}  
     \multirow{5}{*}{\rotatebox{90}{Desired Behavior}} & 

    \begin{tabular}{@{}l@{}} SDB1. Define properties using a\\logic-based language \end{tabular} & \cite{luckcuck:2019} & \cite{santos:2021,hu:2019,yamaguchi2023rtamt} & \cite{santos2021safety,menghi:2021,aldegheri2021containerized,lesire:2019}  & \begin{tabular}{@{}l@{}}  MTL~\cite{santos:2021,hu:2019} \\ rtamt4ros~\cite{nickovicrtamt}\href{https://github.com/nickovic/rtamt4ros}{\faicon{github}} \\ LTL~\cite{aldegheri2021containerized} \\ Past-Time LTL~\cite{lesire:2019}  \end{tabular}   \\\cline{2-6}
    
     & \begin{tabular}{@{}l@{}} SDB2. Use Domain Specific\\ Languages (DSLs) to specify\\ properties \end{tabular} & \cite{luckcuck:2019} & \begin{tabular}{@{}l@{}}\cite{huang:2014,ferrando:2020,stadler2023romosu} \\ \cite{kirca2023runtime,degirmenci2023developing}
     \end{tabular}& \cite{adam2016rule}  & \begin{tabular}{@{}l@{}} ROSRV~\cite{huang:2014}\href{https://github.com/cansuerdogan/ROSRV}{\faicon{github}} \\ ROSMonitoring~\cite{ferrando:2020}\href{https://github.com/autonomy-and-verification-uol/ROSMonitoring}{\faicon{github}}  \\ RuBaSS~\cite{adam2016rule} \end{tabular}
     \\\cline{2-6}

    & \begin{tabular}{@{}l@{}} SDB3. Use languages and tools to\\scenario-based specification of\\ test cases \end{tabular} & \cite{afzal:2020,luckcuck:2019,bertolino:2021}& \cite{balakrishnan:2021,duan2021implementing,liu:2017} & \cite{queiroz2022driver,fremont2022scenic,xu2022sil}  & \begin{tabular}{@{}l@{}} Geoscenario~\cite{queiroz2022driver}\href{https://github.com/rodrigoqueiroz/geoscenarioserver}{\faicon{github}} \\
     SCENIC~\cite{fremont2022scenic}\href{https://github.com/BerkeleyLearnVerify/Scenic}{\faicon{github}} \\
     PerceMon~\cite{balakrishnan:2021}\href{https://github.com/CPS-VIDA/PerceMon.git}{\faicon{github}} \\
     pedsim\_ros~\cite{liu:2017}\href{https://github.com/srl-freiburg/pedsim_ros}{\faicon{github}} \\ 
     \end{tabular} \\
     

    \midrule
    \midrule
    
    \multirow{1}{*}{\rotatebox{90}{Monitor and}} 
    \multirow{1}{*}{\rotatebox{90}{Test Automation}} & 

    \begin{tabular}{@{}l@{}} MTA1. Improve the robustness of  \\ the system by performing noise  \\and fault injection \end{tabular} & \cite{bertolino:2021} & \cite{hutchison2018,bai2024multi, xie2022rozz}  & \cite{yayan2023,hsiao2021mavfi,10.1145/3540250.3549164} & \begin{tabular}{@{}l@{}} ros1\_fuzzer\href{https://github.com/aliasrobotics/ros1_fuzzer}{\faicon{github}} \\
    ros2\_fuzzer\href{https://github.com/aliasrobotics/ros2_fuzzer}{\faicon{github}} \\ RoboFuzz~\cite{10.1145/3540250.3549164}\href{https://github.com/sslab-gatech/RoboFuzz}{\faicon{github}} \\
    imfit~\cite{yayan2023}\href{https://github.com/inomuh/imfit}{\faicon{github}} \\ camfitool\href{http://wiki.ros.org/camfitool}{\faicon{github}}\\ROSPenTo\cite{dieber2020penetration}\href{https://github.com/jr-robotics/ROSPenTo}{\faicon{github}}\end{tabular}  \\ \cline{2-6}
     
     &\begin{tabular}{@{}l@{}} MTA2. Exploit automation for test \\case generation, test case \\prioritization and selection, oracle  \\and monitor generation\end{tabular} & \cite{bertolino:2021,afzal:2020} & \cite{hutchison2018} & \cite{ortega2022testing}  & \begin{tabular}{@{}l@{}} 
     \vspace{.1cm} Equiv. Part.\cite{ortega2022testing}\\ 
     Mithra\cite{afzal2021mithra}\\ 
     Trajectory Generator\cite{hildebrandt2020feasible}\\
     Hypothesis~\cite{santos2018property}\href{https://github.com/HypothesisWorks/hypothesis}{\faicon{github}} 
     \end{tabular} \\ 

    \midrule
    \midrule

    \multirow{1}{*}{\rotatebox{90}{System}} 
    \multirow{1}{*}{\rotatebox{90}{Execution}} 

     &\begin{tabular}{@{}l@{}} SE1. Use record-and-replay when\\performing exploratory field tests \end{tabular} & \cite{bertolino:2021,afzal:2020} & \cite{beul:2017} &  \begin{tabular}{@{}l@{}}\cite{han2022mixed,ortega2022testing,zahn2019optimization} \\ \cite{Caldas2020,Gil2021} \end{tabular} & \begin{tabular}{@{}l@{}} rosbag~\href{https://github.com/ros/ros_comm/tree/noetic-devel/tools/rosbag}{\faicon{github}} \\ rosbag2~\href{https://github.com/ros2/rosbag2}{\faicon{github}} \\ Rerun.io~\href{https://github.com/rerun-io/rerun}{\faicon{github}} \\  NUbots~\cite{zahn2019optimization}\href{https://github.com/NUbots/NUbots}{\faicon{github}}\\ BSN~\cite{Caldas2020,Gil2021}\href{https://github.com/lesunb/bsn}{\faicon{github}} \end{tabular} \\\cline{2-6}
     
     & \begin{tabular}{@{}l@{}} SE2. No GUIs! Prioritize headless\\simulation \end{tabular} & \cite{bertolino:2021, afzal:2020} & \cite{hutchison2018} & \cite{afzal2021simulation,timperley2018crashing}  & \begin{tabular}{@{}l@{}}
     CI \cite{estivill2018continuous} \\ 
     OpenDaVinci~\cite{berger2016open}\href{https://github.com/se-research/OpenDaVINCI/tree/master/automotive/miniature/studies/example-boxparker}{\faicon{github}}
     \end{tabular}\\

    \midrule
    \midrule
    
    \multirow{1}{*}{\rotatebox{90}{Analysis \&}} 
    \multirow{2}{*}{\rotatebox{90}{Reporting}} &

    \begin{tabular}{@{}l@{}} AR1. Perform postmortem analysis\\to diagnose non-passing test cases. \end{tabular} & \cite{falcone:2021} & \cite{roman2018overseer} & \cite{kirchner2013rosha,hossen2023care,morais2015distributed}  & \begin{tabular}{@{}l@{}} Overseer\cite{roman2018overseer} \\
RoSHA~\cite{kirchner2013rosha} \\ CARE~\cite{hossen2023care}\href{https://github.com/softsys4ai/care}{\faicon{github}} \\ Rason\cite{morais2015distributed}\href{https://github.com/lsa-pucrs/rason/}{\faicon{github}}\end{tabular} \\ \cline{2-6}
    & \begin{tabular}{@{}l@{}} AR2. Use reliable tooling in order\\to manage field data\end{tabular} & \cite{bertolino:2021,malavolta2021mining} & \cite{lahami2016safe,hartsell:2019,roman2018overseer} & \cite{ortega2022testing,tampier2022field}  & \begin{tabular}{@{}l@{}}wstool\footnote{Named also rosws, and vcstool.}~\cite{ortega2022testing}\href{http://wiki.ros.org/wstool}{\faicon{github}} \\ Overseer~\cite{roman2018overseer} \\ FTT~\cite{tampier2022field}\href{https://github.com/fkie/field_test_tool}{\faicon{github}} \\
    warehouse\_ros\href{https://github.com/ros-planning/warehouse_ros_mongo}{\faicon{github}} \end{tabular} \\ 



    \bottomrule
    \end{tabular}%
    \label{tab:guid_sources2}
\end{table*}

\noindent{{\bf Process (HOW)}} We divide, according to Falcone et al.~\cite{falcone:2021}, techniques for gathering information in two: inline, and outline. Inline logging and filtering stand for techniques that ask for actually inserting snippets of code in the system under scrutiny as a means to provide an API for logging, e.g.,~\cite{kaveti2021ros}. Outline logging and filtering stands for techniques that enable an external means to gather and filter information that does not require changing the source code, e.g.,~\cite{ferrando:2020,elbaum2000software,mitrevski2022deploying}. Developers may choose one or another given their domain of application.

\looseness=-1
\noindent{{\bf Exemplars}} Exemplar implementations for providing an API for logging and filtering may be inline or outline. 

\looseness=-1
\noindent\emph{Inline Logging and Filtering.} ROS, by standard, contains a system-wide logging mechanism, namely ROS logging (\href{http://wiki.ros.org/roscpp/Overview/Logging}{wiki: logging}). ROS logging works with macros for instrumenting the ROS nodes with information hooks. In the background, the macros send messages with the information to be logged through a standard topic called rosout. On the other side of the topic, an extra node, within the roscore package, persists the data in a textual format (\href{https://github.com/ros/ros_comm/blob/noetic-devel/clients/rospy/src/rospy/impl/rosout.py}{git: noetic-devel/ros/../rosout~\faicon{github}}). Developers can use the macros to log information that might be used for testing in a later stage, given the application-specific requirements. For instance, the Amazon AWS service for robotics provides AWS CloudWatch Logs (\href{https://github.com/aws-robotics/cloudwatchlogs-ros2}{git: aws-robotics/cloudwatchlogs-ros2~\faicon{github}}) interfaces directly with rosout to monitor applications using the standard ROS logging interface. In addition, the standard library provides logging macros with embedded filtering capabilities, which enables eliminating noise from the logged events and can render a useful tool for testers.
ROS Rescue~\cite{kaveti2021ros} is another example of inline logging. The tool aims to solve the ROSMaster problem as a single point of failure. The authors approach check-pointing and restoring state by logging changes in the metadata stored in the master node. Such metadata contains URIs from various nodes, port numbers, published or subscribed topics, services, and parameters from the parameter server. Kaveti et al. create an API for the ROS master node (\href{https://github.com/PushyamiKaveti/fault-tolerant-ros-master/blob/master/src/ros_comm/rosmaster/src/rosmaster/master_api.py}{git: master\_api.py~\faicon{github}}) using the official logging library from Python
to persist metadata in YAML format. Their technique opens space for further inspection of ROS applications without access to the source code.\\\vspace{1mm}

\emph{Outline Logging and Filtering.} ROSMonitoring~\cite{ferrando:2020} employs monitors to persist events in textual format. The monitors contain filtering capabilities to eliminate entries that are inconsistent with the specification (requires an oracle). In this context, the launch files to configure ROSMonitoring can be seen as an API for logging and filtering (\href{https://github.com/autonomy-and-verification-uol/ROSMonitoring/blob/master/generator/online_config.yaml}{git: ROSMonitoring/../online\_config.yaml~\faicon{github}}). Monitors in ROSMonitoring are nodes, so the API is a set of known topics and message formats. The tester, in that case, only needs to specify what topics ROSMonitoring will listen to and the type of message to be recorded. Logging and filtering happens within a separate service. On a similar stance, and inspired by aircraft black box (and software black box~\cite{elbaum2000software}), Mitrevski et al.~\cite{mitrevski2022deploying} proposes the concept of Robotic Black Box (\href{https://github.com/ropod-project/black-box}{git: ropod-project/black-box~\faicon{github}}). The black box operates as an isolated component responsible for listening to data traffic from distinct sources and logging in an easily retrievable medium. Similarly to ROSMonitoring, Mitrevski's black box approach to logging (\href{https://github.com/ropod-project/black-box/blob/master/pybb/logger_main.py}{git: ropod-project/black-box/../logger\_main.*~\faicon{github}}) inspects topics and message types that are configured in advance. Different from ROSMonitoring, Robotic Black Box offers logging not only in textual format but also in a MongoDB database, which supports data processing, filtering, and retrieval. Robotic Black Box stands out when it comes to its filtering and retrieval capabilities. The approach builds a customized query interface over the MongoDB database using names of collections, timestamps and metadata to filter the results (\href{https://github.com/ropod-project/black-box-tools/blob/master/black_box_tools/db_utils.py}{git: black\_box\_tools/db\_utils.py~\faicon{github}}).

\noindent{{\bf Strengths}} An API for retrieval and filtering facilitates access to valuable information resulting in effortless observability of inner states and events. 

\noindent{{\bf Weaknesses}} Overuse of logging may result in performance issues. Incorrectly implemented logging and filtering capabilities may lead to noise in the data, impacting the reliability of the tests. Insufficient logging may result in an incomplete assessment of the system behavior and might generate false positives. Finally, it may also lead to security breaches if not done with caution.
\vspace{-0.2cm}
\subsection{Guidelines for Quality Assurance through Runtime Verification and Field-Based Testing}\label{sec:guide_testing}

We synthesized the guidelines in Tab.~\ref{tab:guid_sources2} relevant to providing runtime verification and field-based testing quality assurance through \textit{testing} activities. Analogous to the guidelines presented in \cref{sec:guide_facilitate}, the guidelines for provisioning quality assurance through \rvft~are organized in groups and summarized in Fig.~\ref{fig:dev-test-process}). We trace each guideline to the sources of information, namely, the relevant initial surveys (column Initial), the systematic literature review (column Lit. Review), the specific search phase (column Specific Search), and the repositories mining (column Exemplars). 
The Prepare Execution Environment (PE) group contains two guidelines: PE1 to warn about the overhead acceptance criteria, and PE2 to create models for runtime assessment. The Specify (Un)-Desired Behavior group (SDB) contains three guidelines concerning the specification of desired and/or undesired behaviors. SDB1 and SDB2 concern the specification of properties through the use of logic-based languages (SDB1) or Domain Specific Languages (DSLs) (SDB2). SDB3, in turn, focuses on scenario-based specifications of test cases. 
The  Generate Monitors and Test Cases group (MTA) contains two guidelines. MTA1 focuses on how to improve the robustness of the system by performing noise and fault injection. MTA2 discusses how to exploit automation for monitoring and testing, e.g., generation and prioritization of test cases. The System Execution group (SE) contains two guidelines focusing on the importance of using record-and-replay when performing exploratory field tests (SE1) and the importance of headless simulation (without GUI) for optimization and/or automation (SE2). Finally, the Analysis \& Reporting group (AR) contains two guidelines. AR1 focuses on performing postmortem analysis to diagnose non-passing test cases. While AR2 focuses on the use of reliable tooling to manage field data.
In the remainder of this section, we show in detail the guideline PE1 on understanding the overhead acceptance criteria. 

We provide an in-depth view of a representative guideline (PE1) for understanding the overhead acceptance criteria in the preparation of the execution environment.

\vspace{-0.2cm}
\subsubsection*{PE1. Understand the overhead acceptance criteria}\label{sec:PE1}
        
    The guideline PE1 recommends quality assurance teams understand the monitoring, isolation, or security and privacy overhead acceptance criteria as a basis for deciding the runtime assurance strategy given that performance is fundamental in robotics applications. Using PE1 helps to avoid unexpected interruptions in the robotic mission due to the testing apparatus. However, the overhead calculation process may turn laborious and conflict with time-to-market requirements. 

\vspace{.2cm}
\noindent{{\bf ID}} PE1

\noindent{{\bf Title}} Understand the overhead acceptance criteria

\noindent{{\bf Context (WHEN)}} According to practitioners, performance is among the three most important quality attributes when designing a robotic application in ROS~\cite{malavolta2021mining}, intended as the degree to which a robotic application performs its functions within specified time and is efficient in the use of resources. Performance is often important in robotics because many computations performed by robots tend to be data-intensive, e.g., computer vision, planning, and navigation~\cite{malavolta2021mining}. Gaining confidence in, e.g, correctness, reliability, robustness, of robotic applications, then, should not interfere with the nominal performance of the robot under scrutiny. Less (or no) impact on performance is especially desirable when the assurance gathering process interacts with the running system, for instance, in-the-field testing and runtime verification. 
        We name the extra load available for gaining confidence on ROS-based applications as \textit{overhead acceptance criteria}. The overhead acceptance criteria can be allocated to monitoring,  isolation, or maintaining the security of privacy during the runtime assessment session~\cite{bertolino:2021}.
            
\noindent{{\bf Reason (WHY)}} A rule of thumb says that more observations tend to enable more precise assurance arguments within a limit. 
        Observing, however, is never free from side effects, incurring in overhead~\cite{falcone:2021}. Therefore, the quality assurance team must contrast the precision of the assurance arguments against overhead introduced by the observation medium. The overhead acceptance criteria are the basis for deciding the runtime assurance strategy. 
            
\noindent{{\bf Suggestion (WHAT)}} 
        The use of runtime verification or field-based techniques might add computational overhead.
        The QA team should understand how much overhead is acceptable; this is important to decide on a test strategy that will not severely impact the performance of the running system. Such overhead may be due to monitoring with ros\_tracing (\href{https://github.com/ros2/ros2_tracing}{git: ros2/ros2$\_$tracing~\faicon{github}}), component isolation, or security and privacy maintenance overhead with ROSploit (git: seanrivera/rosploit).
            
\noindent{{\bf Process (HOW)}} Typically, understanding the overhead acceptance criteria follows from contrasting the required performance for delivering the required service, aka nominal performance (e.g., time to reaction, latency, speed), against available resources (e.g., computing power). The QA team may use off-the-shelf ROS tooling, like \href{https://gitlab.com/ApexAI/performance_test}{git: ApexAI/Performance$\_$test~\faicon{gitlab}}, to understand the nominal performance of the ROS-based application. The difference between nominal performance and expected performance may be allocated to the overhead acceptance criteria. If the nominal performance is close enough to the expected performance, there is not enough space for implementing runtime assurance techniques. We categorize three types of overhead that may affect the runtime assurance process: Runtime Monitoring~\cite{ferrando:2020,BEDARD2023104361}, Isolation overhead~\cite{lahami2016safe,silva2022self}, and Security and Privacy overhead~\cite{rivera:2021,breiling2017secure}.

\noindent{{\bf Exemplars}} Exemplar overhead analyses may be due to monitoring, isolation, and security and privacy. 
        \emph{Monitoring Overhead} is all extra load put on the system-under-test due to gathering, interpreting, and elaborating data about the execution. For example, ROSMonitoring (\href{https://github.com/autonomy-and-verification-uol/ROSMonitoring}{git: autonomy-and-verification-uol/ROSMonitoring~\faicon{github}})~\cite{ferrando:2020} determines the monitoring overhead by calculating the delay introduced in the message delivery time between ROS nodes. The authors analyze the overhead by varying the size of the system under monitoring, message passing frequency, and number of monitor nodes. However, their overhead analysis is not transparent for gathering, interpreting, or elaborating on data, the analysis looks at the monitoring overhead as a black box. Another example is ros2 tracing (\href{https://github.com/ros2/ros2_tracing}{git: ros2/ros2$\_$tracing~\faicon{github}})~\cite{BEDARD2023104361} that provides a tool for tracing ROS2 systems with low latency overhead. The authors measure monitoring overhead by collecting the time between publishing a message and when it is handled by the subscription callback. In short, monitoring and tracing tools add some overhead that typically affects the message passing latency, the QA team must understand and define a precise time allowance for this overhead.\\
        \emph{Isolation Overhead} is all extra load put on the system-under-test to guarantee that the runtime assessment will not interfere with the normal operation or produce undesired side effects. As an example, Lahami et al.~\cite{lahami2016safe} propose a safe and resource-aware approach to test dynamic and distributed systems. 
        They achieve safety by employing testing isolation techniques such as BIT-based, tagging-based, aspect-based, cloning-based, and blocking-based and resource awareness by setting resource monitors such as processor load, main memory, and network bandwidth.
        The authors show that the overall overhead is relatively low and tolerable, mainly if dynamic adaptations are not commonly requested. However, there is no fine-grained evaluation of the overhead introduced by isolation techniques. In fact, isolation overhead is rarely reported~\cite{silva2022self}.\\
        \emph{Security and Privacy Overhead} is all extra load put on the system-under-test for maintaining security and privacy constraints while testing. For example, ROS-Immunity is a security tool for defending ROS-based applications from malicious attackers. Rivera et al.~\cite{rivera:2021}, determine overhead for maintaining the ROS-based system secure, while operating, in terms of power consumption (in Watts) by comparing the system's power draw with and without their tool. From another stance,  Breiling et al~\cite{breiling2017secure} present a secure communication channel to enable communication between ROS nodes using protocols such as Transport Layer Security (TLS) and Datagram Transport Layer Security (DTLS) per each ROS topic in the application. The protocols follow three steps: an initial handshake with mutual authentication, using symmetric encryption (AES-256), and using Message Authentication Codes (MAC) for data integrity. Importantly, they evaluate the overhead introduced for each step, which amounted to a few percentage points (2\%-5\%) of increase in average CPU load.

\noindent{{\bf Strengths}} Understanding the overhead acceptance criteria in advance may avoid unexpected interruptions in the ROS application functionality due to runtime quality assessment procedures. For example, when testing procedures are mistakenly scheduled whenever the ROS-based application is operating under a high load. Learning about the overhead in advance criteria may also ask for re-design due to a lack of verifiability during runtime. For example, when the nominal performance of the ROS system is close to the expected performance, in value.
            
\noindent{{\bf Weaknesses}} Although there is tooling to support the assessment of the overhead acceptance criteria, the process for understanding involves executing the system, collecting performance data, and analysis. This may conflict with time-to-market requirements, asking for further negotiation with the business goals of the ROS-based system.
\section{Validation}\label{sec:results}

This section discusses the outcomes of the validation with experts from industry and academia made through questionnaires (available in the replication package section of the website~\cite{guidelines}). 
Our goal is to check whether the guidelines synthesized from 
literature and insights extracted from open-source ROS repositories are considered to be useful, clear, and applicable by developers and QA teams testing and verifying ROS code.
To this end, we elicit three hypotheses per guideline:

\begin{itemize}
    \item[$H_1.$] Overall, the guideline is useful. {\it (Usefulness)}
    \item[$H_2.$] The formulation of the guideline is clear. {\it (Clarity)}
    \item[$H_3.$] The guideline is applicable to ROS-based systems. {\it (Applicability)} 
\end{itemize}

Differently from usefulness, we note that applicability refers to the extent to which the respondents consider that the guideline could be directly applied to ROS-based systems that {\it they have worked with} at the moment of the survey or in the past. In turn, usefulness refers to ROS-based systems in general.

It is important to note that applicability, clarity, and usefulness are not independent aspects. In~\cref{sec:discussion} we will investigate discrepancies among them to identify research gaps and research challenges for future work.

\subsection{Participants}

We received 55 answers to the questionnaire, of which 33 were from developers and 22 from quality assurance (QA) teams. To obtain those answers, we sent 1032 emails out of which: 306 targeted paper authors, 772 targeted developers (extracted from repositories), and 68 targeted experts we met personally. We excluded 114 emails due to intersections, e.g. paper authors that were also targeted developers. Additionally, we shared posts in the ROS discourse forum. Duplicated emails were removed. The 39-day campaign stopped when we did not receive any more answers from the candidate respondents. The questionnaires were tailored to check the three hypotheses. In addition, we asked each respondent to answer four other questions: (i) for how long they have worked in robotics, (ii) what their experience with ROS was, (iii) what kinds of organizations they have worked in, and (iv) what robotics domains they have worked on. 
Based on their answers, we created a summary of five fictitious but representative profiles, as described in Table~\ref{tab:demographics}.   

\begin{table*}[htb!]
    \centering
    \begin{tabular}{ccllllc}
        \toprule
        {\bf ID}  & {\bf Time}  & {\bf Experience} & {\bf Organization} & {\bf Domain} & {\bf Role} & {\bf\# Individuals} \\
        \midrule
        P1 & $>$10y & Contributed to ROS packages  & Academia and Industry       & Service Robotics      & Developer & 18\\
        P2 & 3--5y  & Worked on app. using ROS     & Industry                    & Industrial automation & Developer & 17\\
        P3 & 1--3y  & Contributed to ROS packages  & Academia                    & General-purpose       & Developer & 6 \\
        P4 & 1--3y  & Worked on app. using ROS & Academia and Industry           & General-purpose       & QA        & 3 \\
        P5 & 3--5y  & Worked on app. using ROS & Academia and Independent Groups & Marine Robotics       & QA        & 11\\
        \bottomrule
    \end{tabular}
    \caption{Summary of the representative profiles for the 55 questionnaire respondents. Note: app. stands for applications}
    \label{tab:demographics}
\end{table*}

\subsection{Questionnaire Results}

The Likert plot in Fig.~\ref{fig:likert} provides an overview of the answers from the questionnaire. While the majority of votes lean to the `Agree' side, some guideline descriptions (i.e., CD2, CI1, CI2, CI3, I1, I2, I3, I4, MTA1, MTA2, PE2, SDB1, SDB2, SDB3, SE2) received `Strongly Disagree' votes. This subsection mostly analyses disagreements in contrast to agreements since they might shed light on new research opportunities. To clarify the disagreements, we collected comments left in free text from the online questionnaire\footnote{We make available the anonymized transcript of all comments in the replication package~\cite{guidelines}.}. The questionnaire targeted per group comments, in which respondents could direct their suggestion either to the whole group or to specific guidelines, usually indicating their identifier.

Guideline CD2 concerns ensuring global time monotonicity of events and states to address the potential non-determinism in the scheduling of events in ROS-based applications. 
In general, comments left to CD2 emphasize the need to ensure the monotonicity of events, {\it ``While typically small, occasionally you will see a large time jump if NTP does something unexpected. With enough robots and enough usage, these (and surely other) odd events will occur.''}. While some comments question whether monotonicity of events should be an explicit or implicit property of the design phase, there is a suggestion on how to address the time synchronization problem using the NTP/PTP time protocols, i.e.,{\it ``NTP/PTP status and verifying that these time protocols have converged to a stable solution with small offsets''}.

Guideline CI2 concerns security and privacy, which are often under-explored in the design
of ROS-based applications. Threats to privacy and security may be catastrophic and may be an unacceptable side-effect of runtime assessment of such robotic applications. 
During the validation of this guideline, the respondents were mostly unsatisfied with the wording. They claim that the guideline uses {\it ``complicated language''} or {\it ``worded confusingly''}. Ultimately asking for a guideline refinement with a better understanding of the SROS tool and the Alias Robotics company, and leveraging separate criteria to form security constraints: {\it ``(1) software BOM and patch process, (2) application security, and (3) user data privacy controls''}.

Guideline I3 concerns providing an API for stimulating unexpected scenarios for
testing the robustness of the target system; this can help addressing
challenges in finding both fault and error that are representative
of real software faults.
During the validation of this guideline, respondents claim that they agree with the guideline's intention but warn that implementing such a solution might lead to little return on investment, they assert that {\it ``[...] it will result in a lot of development to implement fake faults, and you're still going to miss so many real faults''}. While they say that {\it ``Building tools to simulate faults can also be very difficult and time-consuming.''}, they also highlight that such approaches also introduce safety risks when injecting faults in field tests: {\it ``Injecting faults as part of the codebase in safety systems [...] would never be approved by certification entities. [...].''}.

Guideline SDB1 concerns the use of a logic-based language for specifying properties that 
describe observable actions, outputs, how they relate to each other and when they
should manifest. To simplify the specification of properties in temporal logic, some of the
tools refer to existent property specification patterns~\cite{patterns2022}, which also enable the formulation of properties in structured and unambiguous English. 
Guideline SDB2 proposes to address the specification problem with a code-like alternative, a built-in ROS-tailored language allowing quality assurance teams to specify and validate the correct behaviour of the system. Focusing on testing, SDB3 proposes to use scenario-based test case generation approaches for the systematic exploration of different situations and conditions that the robotic system may encounter in the real world.
For the SDB group, the respondents discussed that they believe that behavior specification is an activity that is targeted to senior developers rather than QA experts. They say {\it ``I wish QA teams could do such advanced tasks. Those are senior developers, not QA''}.

In light of this discussion, we have some understanding concerning the applicability of the SDB group of guidelines, as well as of CD2, CI2, and I3. 

Guideline I1 recommends the use of ROS nodes with lifecycle management to provide a structured way to manage the state of the nodes and the interactions between them.
This structure helps (i) ensuring that nodes are in the right
state for testing, (ii) guaranteeing that the interactions between nodes
are predictable, and (iii) mitigating dangling references to nodes that are no longer in use. Concerning I1, this guideline recommends an API for querying and updating internal lifecycle. This is considered to be, overall a useful guideline, but we understand that it requires special conditions to realize it, e.g., full control of the robotic system. It is interesting to highlight that this guideline is very similar to guideline B1 in~\cite{malavolta2021mining}, where the authors suggest that the behavior of each node should follow a well-defined life cycle, which should be queryable and updatable at runtime. 

We do not have enlightening comments for the remaining guidelines CI1, CI3, I2, I4, MTA1, MTA2, PE2, and SE2. However, the overall evaluation of many of them is quite positive and no further investigation is needed.
Instead, we feel that it would be worth better investigating the usefulness, clarity, and applicability of guidelines MTA2, PE2, and CI2, which are those with scores of applicability lower than 65\% and for which we have no further information from the questionnaires. To better analyze the validation results we performed statistical and practical significance analysis.



 

\subsection{Statistical Analysis}

In complement to the Likert plot analysis, we back up the responses using statistical analysis. First, we assume that the answers from the Likert items are symmetric and this allows us to use boxplots and visualize whether the hypotheses ($H_{1} - H_{3}$) hold for each guideline from a bird's eye view. We map the Likert items to integer values according to the following rule:  {\it Strongly~Disagree}=1, {\it Disagree}=2, {\it Agree}=3, {\it Strongly~Agree}=4. The resulting Fig.~\ref{fig:boxplot} compares the boxplots for each guideline (y-axis) for the attributes we tested against, i.e., applicability, clarity, and usefulness; while the x-axis determines the Likert Items, i.e., Strongly Disagree, Disagree, Agree, and Strongly Agree.

\begin{figure*}[htb!]
    
    \begin{subfigure}{\columnwidth}
      \centering
      \includegraphics[width=\columnwidth]{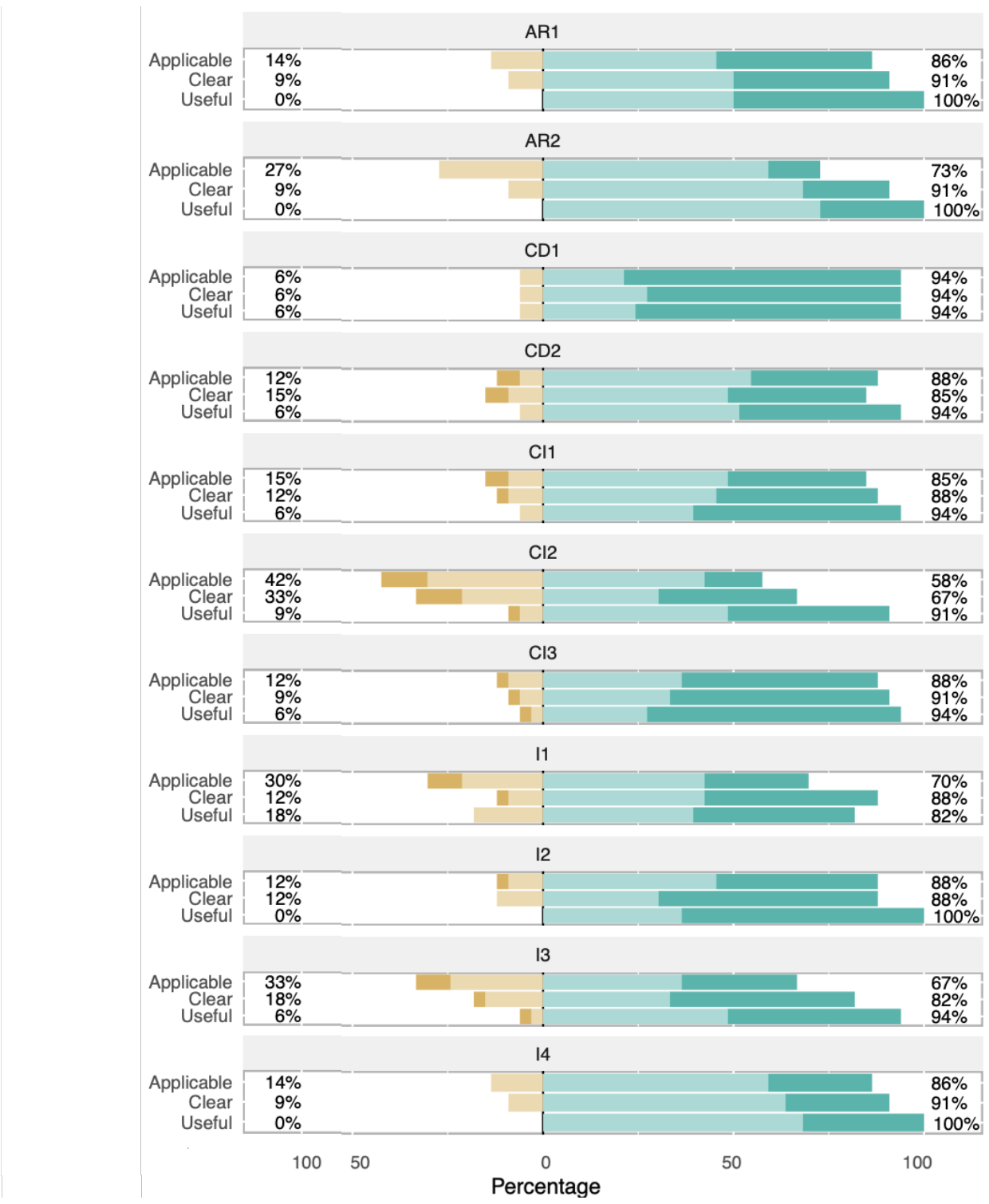}
      \label{fig:likert_a}
    \end{subfigure}
\begin{subfigure}{\columnwidth}    
    \begin{subfigure}{\columnwidth}
      \centering
      \includegraphics[width=\columnwidth]{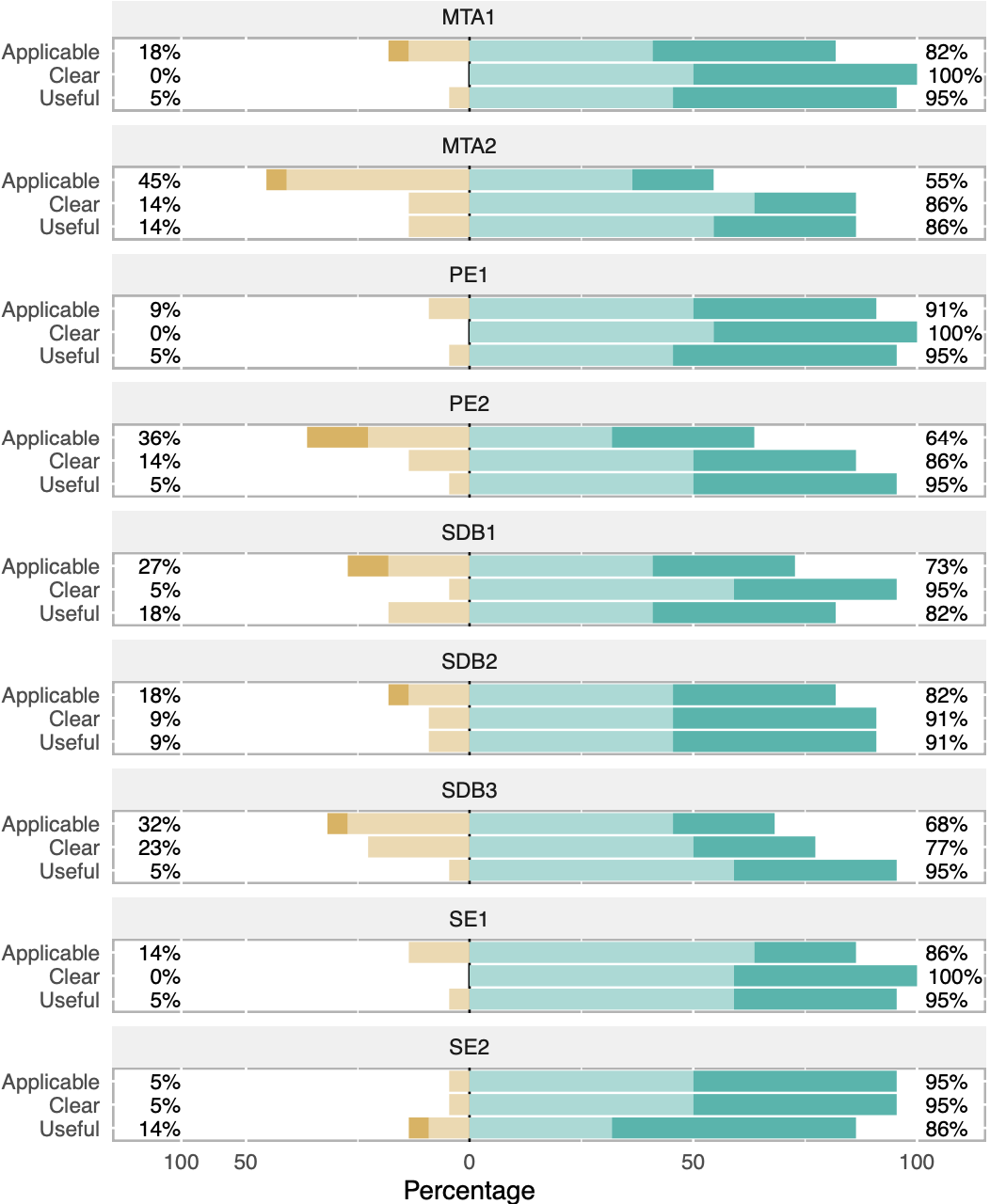}
      \label{fig:likert_b}
    \end{subfigure}

    \begin{subfigure}{\columnwidth}
      \centering
      \includegraphics[width=\columnwidth]{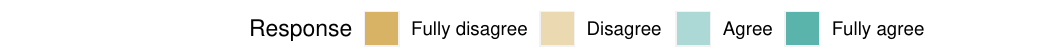}
      \label{fig:legend_likert}
    \end{subfigure}
     \end{subfigure}   
    \caption{Likert plot for the questionnaire answers}
\label{fig:likert}
\end{figure*}

\begin{figure*}[htb!]
    \centering
    \includegraphics[width=0.8\textwidth]{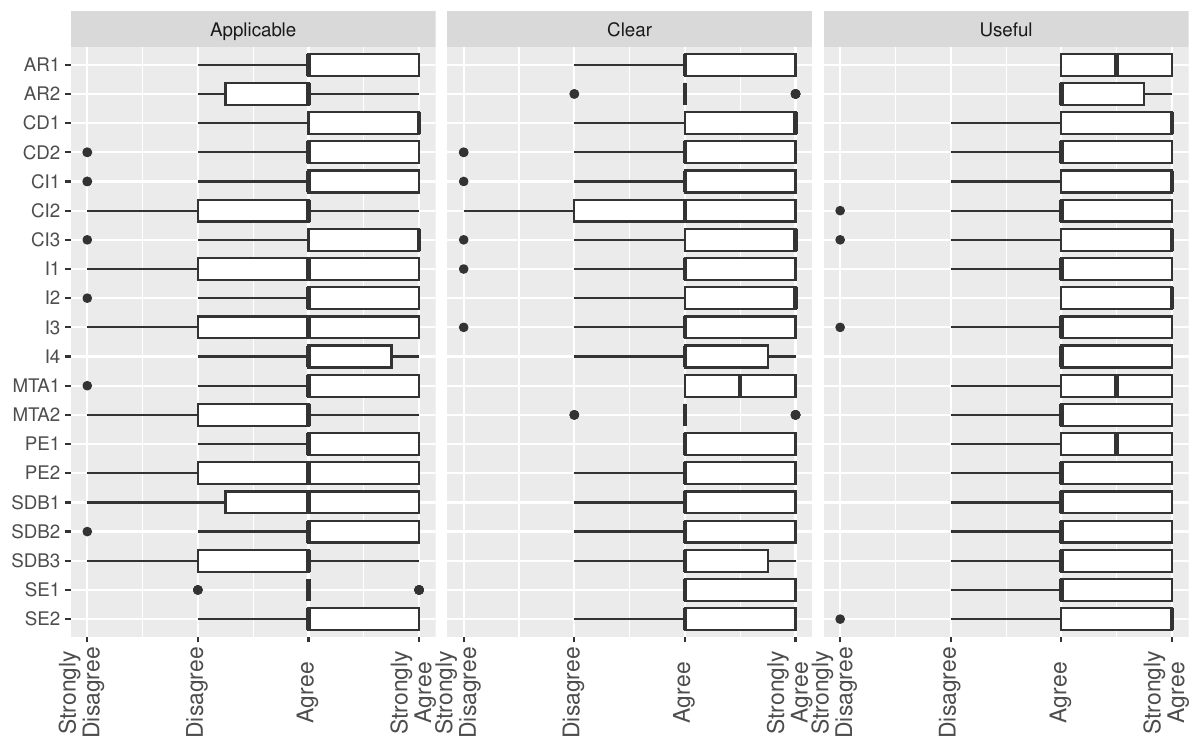}
    \caption{Boxplots for the results bird's eye view\vspace{2mm}}
    \label{fig:boxplot}
\end{figure*}

 At first glance at the plot from Fig. \ref{fig:boxplot}, we observe that the {\it medians} revolve around {\it Agree} and {\it Strongly Agree} for the three dimensions (i.e., applicability, clarity, and usefulness). Therefore, overall developers and QA teams agree that the synthesized guidelines are applicable, useful, and clear.
 
 In light of such results, we further analyzed the statistical and practical significance regarding the agreement for these guidelines (see \cref{sec:significanceAnalysis}). For those guidelines where the significance test failed or practical significance had a small effect, we conducted a follow-up discussion with the questionnaire respondents to understand the causes for disagreement (see \cref{sec:followup}).

\subsubsection{Significance Analysis}\label{sec:significanceAnalysis}

In light of the significance analysis, we can claim that all guidelines are {\it applicable}, {\it clearly} formulated, and {\it useful} according to the 55 responses from the questionnaire. We conclude also that guidelines CI2, MTA2, and PE2 ask for more information to draw solid conclusions on whether they are {\it applicable} or not. We reached this conclusion thanks to the analysis detailed in the following.

In preparation for the significance analysis tests, we transform the input data from categorical (i.e., {\it Strongly Disagree}, {\it Disagree}, {\it Agree}, {\it Strongly Agree}), to numeric ({\it Strongly~Disagree~\&~Disagree} = -1 and {\it Strongly~Agree~\&~Agree} = 1). The mapping separates the answers in a format such that both sides entail an equal Euclidean distance from the center, 0. This format allows us to sum the points for each respondent and evaluate whether the aggregate rejects (${\it sum}<0$) or accepts (${\it sum}>0$) the hypotheses $H_1 - H_3$. In addition, we applied Shapiro Wilk's normality test to understand whether the data is normal (available in the validation section of the guidelines website~\cite{guidelines}) and the test reveals that it is not. Thus, we employed the non-parametric one-sample Wilcoxon test for statistical and practical significance.



{\bf Statistical Significance}:
We test our hypotheses for statistical significance by employing the one-sample Wilcoxon test. The test checks whether the sum of all the respondents for each guideline, e.g. AR1, and each dimension, e.g., applicability, is greater than 0 with a 5\% significance level. The resulting plot (Fig.~\ref{fig:wilcoxon}) summarises all the tests, combinations of guidelines, and dimensions.

\begin{figure}[htb!]
    \centering
    \includegraphics[width=.97\columnwidth]{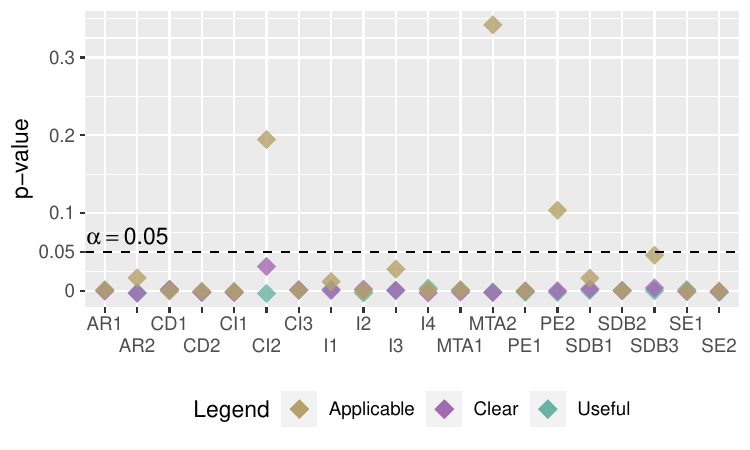}
    \caption{Wilcoxon one-sample test for statistical significance (hypothesis: $\mu \geq 0$)\vspace{2mm}}
    \label{fig:wilcoxon}
\end{figure}

Importantly, Fig.~\ref{fig:wilcoxon} highlights that guidelines CI2, MTA2, and PE2 resulted in p-values larger than 0.05, specifically for {\it applicability}. Precisely, CI2$_{p-value} = 0.1948536$, MTA2$_{p-value} = 0.3416874$, and PE2$_{p-value} = 0.1037148$. The result suggests that the data collected does not confirm that the aforementioned guidelines are applicable to ROS-based systems, as pointed out by the medians from Fig.~\ref{fig:boxplot}. In complement, the tests confirm that the remaining correlation between guideline (below the red dashed line in Fig.~\ref{fig:wilcoxon}) and attributes {\it applicability}, {\it clarity}, and {\it usefulness} is statistically significant.

{\bf Practical Significance}: 
We test our hypotheses for practical significance by calculating the effect size given by the one-sample Wilcoxon signed-rank test. Similarly to the statistical significance test, 
the test checks whether the effect size for one sample Wilcoxon test is greater than zero for the sum of all respondent's answers to the combinations of guidelines and dimensions. The resulting plot (Fig.~\ref{fig:effect_size}) summarises the effect size for all combinations of guidelines and dimensions.

\begin{figure}[htb!]
    \centering
    \includegraphics[width=.97\columnwidth]{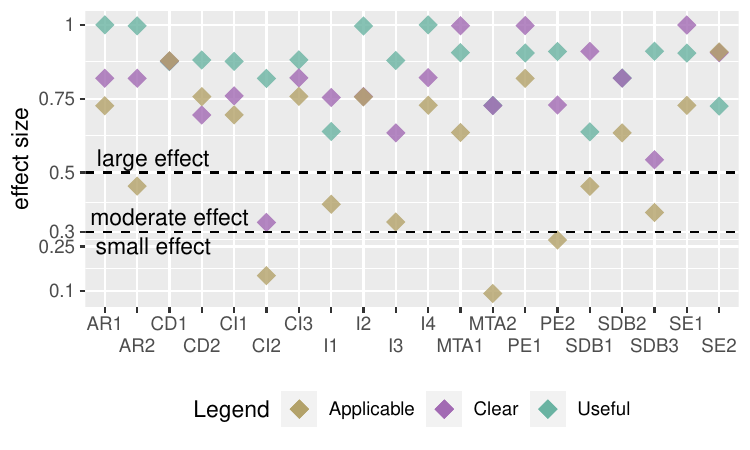}
    \caption{Effect size for one sample Wilcoxon test for practical significance\vspace{2mm}}
    \label{fig:effect_size}
\end{figure}

According to the definitions from Cohen J.~\cite{cohen1992power} limiting values for effect size are small effect $<0.3$, moderate effect $>0.3~{\it and}~<0.5$, and large effect $>0.5$. When it comes to applicability, guidelines CI2, MTA2, and PE2 fall in the small effect category, confirming that we need further data to strengthen our conclusions. Guidelines AR2, CI2, I1, I3, SDB1, and SDB3 are ranked with moderate size effect, which is enough in our case to justify their practical significance. 
Finally, the large-size effect guidelines are the rest, which confirms their practical significance . 

\subsection{Follow-up With Respondents}\label{sec:followup}

Given the lack of statistical and practical significance to guidelines CI2, MTA2, and PE2, we followed up with respondents who willingly left their contact in the questionnaire and opted for disagree or strongly disagree. Then, we sent to the respondents an email asking for further details and reasoning. 

According to their availability to provide further feedback in the follow-up process, we contacted 15 individuals, out of which 10 were academics and 5 were practitioners in the industry. From the 10 responses we received, 7 were academics, and more specifically, 3 were professors, 3 were PhD students and 1 was a technician. Meanwhile, on the practitioner side, out of the 3 responses, two were research engineers and one was a developer/software architect.

In the remainder of this section, we summarize the points made by the researchers and practitioners concerning the applicability of guidelines CI2, MTA2, and PE2 in their line of work.

For MTA2, which refers to exploiting automation tools for test case generation, test case selection, and oracle generation, the main points made were:
\begin{itemize}
    \item As one researcher suggested, the project needs to be large enough to justify developing models for exploiting test and monitor automation: \begin{quote}
    "\textit{If the project is too small, the additional effort required to develop some sort of digital twin is too high compared to the project.}"
\end{quote}
    \item For complex scenarios, the automatic generation of test environments could be computationally too demanding.
\end{itemize}

For PE2, which refers to exploiting system models system or its environment, as well as the creation of digital twins, the main points were:

\begin{itemize}
    \item The amount of work surpasses the project' scale.
    \item It is a difficult challenge to create realistic data for large complex scenarios, even if the model is accurate, given that the computational load can be so high that the real-time factor is so low that it compromises the testing reliability testing. As one expert put it: \begin{quote}
    \textit{``I am working now in a system which the goal is identifying cracks in tunnels, we could not create digital twins (at least yet) because [if] even we have the perfect model, super detailed, of the real world, it is not reliable to run in simulators, because the model is too big, and then the real time factor of the simulation is smaller than 0.3, even with a really good computer.''}
\end{quote}
\end{itemize}

For CI2, which refers to identifying security or privacy vulnerabilities, the major issues raised were:

\begin{itemize}
    \item It is not applicable or relevant to their projects in the current state.
    \item Academic projects usually see no need for such measures as there is not an abundance of sensitive information present.
    \item There are alternative ways to address some of the more frequent vulnerabilities e.g., making the system unreachable from the Internet. As one expert mentions: \begin{quote}
\textit{``We've built a new version of the Gateway around the super-secure Turris Omnia router, the only extra security feature we use from this router is the dynamic firewall Sentinel it offers out of the box.''}
\end{quote}
\end{itemize}

As a result of our follow-up, we revised the guidelines CI2, MTA2, and PE2 for further clarification following the respondents' latest feedback. Acknowledging the experts' comments, we added statements to the {\em Weaknesses} field of the guidelines addressing their concern and better delimiting the scope of applicability each guideline. The modifications are both represented in our online guidelines repository in a pull request\footnote{\href{https://github.com/ros-rvft/ros-rvft.github.io/pull/1/files}{https://github.com/ros-rvft/ros-rvft.github.io/pull/1/files}} and represented in the guidelines in the appendix -- additions are marked with blue highlight and deletions with red.

\section{Discussion}\label{sec:discussion}



We now discuss open challenges in field-based testing~\cite{bertolino:2021} and runtime verification~\cite{falcone:2021,sanchez2019survey}, how our guidelines mitigate them, and what opportunities for future research exist. We selected challenges from the main papers on field-based testing and runtime verification used as initial surveys in our paper. We merged two challenges and excluded two challenges that are not relevant either for the robotics domain or are not focused on software. It is important to highlight that these survey papers do not focus on ROS systems. This explains why some of the challenges are not connected to any guideline, as summarized inTable~\ref{tab:challenges}, which provides a mapping between open challenges and 
guidelines. We will also exploit identified discrepancies among the applicability, clarity, and usefulness validation to identify research gaps and challenges.

\begin{table*}[ht!]
    \centering
    \caption{Open challenges for \rvft of ROS-based systems}
    \begin{tabular}{llc}
    \toprule
    \textbf{Open Challenge} & \textbf{Guidelines} & \textbf{FT or RV?}\\ 
    \midrule
    \textbf{Lack of (Formal) Specifications}~\cite{bertolino:2021}  & \begin{tabular}{@{}l@{}} SDB1, SDB2, SDB3, PE2 \end{tabular} & FT  \\
    \textbf{Generating and implementing field test cases}~\cite{bertolino:2021} 
    & \begin{tabular}{@{}l@{}} MTA2, SEI1, PE2
    \end{tabular} & FT  \\
    \textbf{Isolation Strategies}~\cite{bertolino:2021} 
    & \begin{tabular}{@{}l@{}} I4, CD1, PE1
    \end{tabular} & FT  \\
    \textbf{Oracle Definition}~\cite{bertolino:2021} 
    & \begin{tabular}{@{}l@{}} MTA2
    \end{tabular} & FT  \\
    \textbf{Security and Privacy}~\cite{bertolino:2021,sanchez2019survey} 
    & \begin{tabular}{@{}l@{}} CI2, PE1
    \end{tabular} & RV\&FT  \\
    \textbf{Distributed monitoring}~\cite{falcone:2021,sanchez2019survey}  & \begin{tabular}{@{}l@{}} - 
    \end{tabular} & RV  \\
    \textbf{Monitoring states}~\cite{falcone:2021} 
    & \begin{tabular}{@{}l@{}} I1 \end{tabular} & RV  \\
    \textbf{Richer reactions}~\cite{falcone:2021} 
    & \begin{tabular}{@{}l@{}} - 
    \end{tabular} & RV  \\
    \textbf{Support to imprecise traces}~\cite{falcone:2021} 
    & \begin{tabular}{@{}l@{}} CD2, AR2
    \end{tabular} & RV  \\
    \bottomrule
    \end{tabular}%
    \label{tab:challenges}
\end{table*}

\subsection{Lack of (Formal) Specification} 
Formal specifications are used in test case generation and oracle or monitor definition. However, formal specifications are typically not found in real systems~\cite{bertolino:2021}. This poses a threat to using automated processes to indulging in generating test artifacts. Guidelines on {\it specification of (un)desired behavior} (SDB1, SDB2, SDB3) propose means to discuss and present languages, e.g., LTL, MTL, STL, ROSRV, RuBaSS, Geoscenario, SCENIC, that may assist the specification of behavior. 
Towards this direction, works from Hammoudeh~\cite{garcia2019boot1,hammoudeh2021} take a step forward and propose a model-driven engineering approach by creating meta-models for ROS, further detailed in PE2. What is not defined by their work is the extent to which data from the field may be used in the specifications to shorten the gap between high-level models and concrete real-world scenarios.

According to the evaluation in \cref{sec:results}, these guidelines (SDB ones) have all a discrepancy between usefulness, generally considered as high, and applicability to projects in which respondents have been working, where the evaluation shows less support. This could imply that, even though the respondents see value in this set of guidelines, as supported by the usefulness, the specific conditions of the project in which they have been working were not ideal for these guidelines. This interpretation is supported by the results in~\cref{sec:results}. In fact, respondents reported that behavior specification is an activity that is targeted to senior developers rather than QA teams, which do not have the competencies to do such advanced tasks. The use of languages for the specification of (un)desired behaviors more accessible and easy to use for senior developers (rather than quality assurance experts) would potentially mitigate this discrepancy.


Concerning specification, we can identify a gap between academics and practitioners. In fact, stimulated by research results, we attempted to define a guideline concerning the application of Contract Programming or Design by Contract (DbC), a software design technique that enables the software designer to declare what the code is supposed to do~\cite{Eiffel,reis:2016}, to ROS, where the software components are represented by ROS nodes. 

On the one hand, we found some interesting academic works.
Contracts\_lite~\href{https://github.com/ros-safety/contracts_lite}{\faicon{github}}\footnote{\url{https://github.com/ros-safety/contracts_lite}} is a ROS package from the safety working group that enables the explicit definition of contracts and enforcement of rules. 
As a different alternative, Luckcuck et al.~\cite{luckcuck:2022} take a stance on contract-based verification for ROS systems. In their work, the contracts are manually written using the ROS Contract Language (RCL) and the ROS-based system needs to be abstracted into a model before writing the contracts. 
On the other hand, we did not find much use of DbC for robotics in industrial settings.
We found an informal discussion on this topic on ROS2\footnote{\url{https://discourse.ros.org/t/design-by-contract/2405}}, however, we believe that there are not enough available solutions and examples to formulate it as a guideline.This could be an interesting research direction to investigate.

Still concerning specification, some approaches in space ROS\footnote{\url{https://github.com/space-ros}} use properties implicitly specified. Implicit specification happens when there is a general understanding of a particular desired behavior. For instance, Cobra 
and IKOS 
provide support for two static analysis tools from NASA with embedded or implicit properties. 

We conclude this subsection by highlighting the opportunity to make formal specifications more accessible and usable in practice. A first step in this direction can be found in patterns for properties specification~\cite{autili:2015,menghi:2021,patterns2022}, which we mention in SDB1. This would make it easier for practitioners to use a set of tools for \rvft{} that require a formal specification as input.

\subsection{Generating, Implementing, Orchestrating, and Governing Field Test Cases} 

We proposed various guidelines on field-based testing, as summarized in the following:
\begin{itemize}
\item MTA2 aims at exploiting automation for field-based testing. In this guideline, we describe approaches concerning automatic test case generation and selection.
\item Guideline SE1 focuses on the use of record-and-playback when performing exploratory field tests. Among other recommendations, this guideline suggests using exploratory field testing to find corner cases, and then field endurance tests~\cite{ortega2022testing} to test these corner cases.
\item Guideline PE2 recommends the use of model-based runtime assessment to help manage complexity and ensure the safe operation of ROS-based applications. However, maintaining models always adds overhead.
\end{itemize}

However, we believe that generating, implementing, orchestrating, and governing field test cases includes many other aspects, and test automation in field-based testing is still limited and relies heavily on human contributions as already highlighted in~\cite{bertolino:2021}. In the following we discuss promising future research directions.

\begin{itemize}
\item Because of environment uncertainty, field test cases shall adapt to the operational environment. Also, testers should rely on rules and policies to establish when, how, by whom, and in which order, a selected set of tests can be executed~\cite{bertolino:2021}. This is made clear by the category of testing called self-adaptive testing in the field (SATF)~\cite{SATF}, which focuses on field-based testing approaches that change over time to follow and adapt to the changes and evolution of systems, environments, or users’ behaviors. 
We do not have a specific guideline focusing on this challenge.
\item One of our respondents recommends using state machines when needed. State machines are useful to the extent that the robot knows precisely what state it is in and there is a discrete set of transitions to choose from. However, the robot may not have a single state (e.g., multimodal distribution of pose estimations). Furthermore, the vast majority of important control code pertains to the processes running within states, not to the transition logic between states. So, the recommendation is to not make a state-based architecture and quality assurance system if the majority of bugs do not have anything to do with state transitions.
\item Another respondent would like to see guidelines concerning CI/CD pipelines in the quality assurance process. This can ensure that the code is automatically tested at every stage of development, thus identifying issues early and improving the overall quality of the software.
\item Another of our respondents highlighted the wish for a guideline focusing on the identification of tests that should be re-executed at runtime to avoid the re-execution of all tests at runtime.
\end{itemize}

Because of environmental uncertainty, field test cases shall adapt to the operational environment. Also, testers should rely on rules and policies to establish when, how, by whom, and in which order, a selected set of tests can be executed~\cite{bertolino:2021}.

\subsection{Isolation Strategies} Field test cases must be non-intrusive. They should not intervene with the running processes or their data. However, it can be difficult and expensive to apply strategies to guarantee the isolation of the field test cases in practice~\cite{bertolino:2021}. Guideline I4 focuses on isolating components for testing. Similarly to what was recommended in~\cite{bertolino:2021}, we found that Man-in-the-Middle (MITM) can be a good strategy to provide isolation of computing nodes for field test case isolation.  
However, this can be difficult for robotics. In fact, the two-way
interaction between the software and the physical environment makes it impossible to
assess the behaviour of one in isolation from the other. This can be mitigated, e.g., with model or software in-the-loop testing approaches.

I4 identifies two means to enact MITM in ROS, through a proxy design pattern, e.g., ROSRV~\cite{huang:2014}, or through topic remapping, e.g.,  ROSMonitoring~\cite{ferrando:2020}.
Guideline CD1 recommends that developers strive for ROS nodes with a single responsibility. It should reduce the cost of observing and controlling modules, or skills~\cite{efatmaneshnik2017study}, to facilitate isolation.
Guideline PE1 focuses on understanding the overhead acceptance criteria. It concerns the extra load put on the system-under-test for guaranteeing that the runtime assessment will not interfere with normal operation or produce undesired side effects.

\subsection{Oracle Definition}
Field testing oracles need to adapt to the uncertainty and the unknown execution conditions of the environment~\cite{bertolino:2021}.
We do not have a specific guideline focusing on this challenge. MTA2 aims at exploiting automation for field-based testing, including the oracle generation.
The guideline highlights \textit{Mithra}~\cite{afzal2021mithra}, a novel and unsupervised oracle learning technique for Cyber-Physical Systems (CPS). Further work in this direction would help make field-based testing more popular in robotics. One of the respondents highlights the importance of incorporating guidelines for the use of CI/CD pipelines in the quality assurance process. This would include the use of autonomous oracles to automate the testing phase. 

\subsection{Security and Privacy}
Executing test cases in the field can challenge security and privacy~\cite{bertolino:2021}. This is indeed an important challenge. Guideline CI2 concerns the identification of security and privacy constraints. These constraints can help mitigate this challenge. Guideline PE1 concerns with understanding the overhead acceptance criteria and it focuses also on the overhead due to maintaining security and privacy constraints while testing. 
Besides security and privacy, one of our respondents would like to see guidelines covering broader aspects of quality assurance, including resource management, adaptability, data-driven approaches, system resilience, energy, and ethical considerations. This can be an interesting direction for future research. Integrating other aspects of quality assurance can help in building a more robust, efficient, and reliable quality assurance framework for ROS-based systems, ensuring they are well-prepared for real-world deployment and operation.


\subsection{Distributed Monitoring and Monitoring States}
Decentralization of the monitoring architecture is identified as an area that has not received much attention~\cite{falcone:2021}, probably due to its inherent complexity. We did not identify guidelines related to this aspect.


Concerning monitoring states, it is not very common to find tools able to monitor the states of a program directly~\cite{falcone:2021}. Guideline I1 focuses on APIs for querying and updating the internal lifecycle. The internal states within ROS nodes are typically hidden and not easily accessible. This limits the ability to diagnose and understand unexpected behavior.
Exposing these hidden states is crucial for providing granular information on the system’s behavior, rendering increased observability. Furthermore, managing these hidden states
allows for actions such as starting, stopping, and rolling back to a specific state, in other words, increased controllability.

Most monitoring tools do not support active reactions, like enforcement, recovery, and explanations for declarative specifications~\cite{falcone:2021}. Our respondents 
would like to have guidelines to
design APIs for extra-node enforcement and for bringing a node for a specific state (e.g., rollback). We did not find guidelines related to this aspect.

\subsection{Support to Imprecise Traces}
Runtime verification tools rarely support imprecision in their input traces, e.g., imprecise timestamps, traces with incomplete events or inconsistencies in event sources, or due to purposefully omitted events when sampling~\cite{falcone:2021}. Our guidelines do not provide recommendations or solutions on how to deal with imprecise traces. Instead, some guidelines try to explain how to mitigate imprecision. Guideline CD2 focuses on ensuring global time monotonicity of events and states. Ensuring global time monotonicity of events and states permits addressing the potential non-determinism in the scheduling of events in ROS-based applications. A way to achieve that is to annotate messages with timestamps and synchronization to guarantee ordering.
Also, guideline AR2 promotes the use of reliable tooling to manage field data. The quality assurance team should use reliable tools for field data management to avoid problems with corrupted, unreliable, and/or incomplete data.

\section{Threats to Validity}

{\bf Internal validity} focuses on the level of influence that extraneous variables may have on the design of the study~\cite{wohlin2012experimentation}.
Concerning the systematic literature review, we formalized the design of this study into a detailed research protocol and thoroughly discussed it before the actual execution of the study. 
Our research protocol is based on well-accepted guidelines for systematic literature reviews and mapping studies~\cite{keele2007guidelines}. 
Concerning the repository mining, we strictly followed the approach in ~\cite{malavolta2021mining}, which successfully gathered repositories from ROS. In addition, we make the protocols for the systematic literature review and repository mining available online~\cite{guidelines}. The online material contains all the artifacts generated during the study, including the motivation behind our decisions.

\noindent {\bf External validity} concerns the generalizability of the study's results~\cite{wohlin2012experimentation}.  The generalizability of our study depends strongly on whether the primary studies collected in the literature review represent runtime verification and field-based testing in robotic systems using the robot operating system.
We mitigated the threat by carefully defining the terminology used for the search strings, and inclusion and exclusion criteria. The terminology was validated with researchers credited for their contributions in the domains of robotics (more specifically, in ROS), runtime verification, and field-based testing. Additionally, we use multiple data sources, i.e., ACM Digital Library, IEEE Xplore, Scopus, to increase the coverage of the state-of-the-art. Moreover, we performed an additional round of in-depth searches targeting specific facets of runtime verification and field-based testing in ROS. This covers studies that do not mention runtime verification or field-based testing explicitly, but that can be relevant for the study. 

\noindent {\bf Construct validity} concerns whether the constructs used are suited to answer the research questions~\cite{wohlin2012experimentation}. To mitigate such threats, we performed a thorough quantitative and qualitative validation of each of the guidelines. The validation followed well-established advice surveys in software engineering~\cite{kitchenham2009systematic}. Our study considered experts from different backgrounds within robotics domains and experience with ROS, including a large group of experts who claimed more than 10 years of experience. Whenever the quantitative results rendered inconclusive results, we discussed with experts and presented their comments {\it ipsis litteris} in our paper. The comments are also in the replication package~\cite{guidelines}. 
The questionnaire asked the respondents about usefulness, applicability, and clarity, which might have led to the hypothesis guessing phenomenon~\cite{wohlin2012experimentation}. We mitigate such threat by asking questions specific to each of the guidelines and asking open-ended questions. In light of this, we tried not to make bold conclusions but incorporated the comments, critiques, and suggestions for improvement in the guidelines.

\noindent {\bf Reproducibility} The synthesis of guidelines from diverse sources of knowledge, in our case scientific literature and mined repositories, may turn out to be hard to replicate and reproduce. To mitigate such issues, we provided a guideline specification template that caters to how the information collected from the diverse knowledge source is composed to generate a guideline. 

\section{Related Work}\label{sec:rw}


This section compares our proposed guidelines to approaches for robotics systems, field-based testing, and other literature about devising guidelines for autonomous systems.

{\bf Testing Robotics Systems}. Testing Autonomous and Robotics Systems is an area in expansion. However, it still faces barriers to support the verification and testing~\cite{bozhinoski2019safety,garcia2020robotics}. Song et al.~\cite{song2021concepts}, for instance, elicit testing challenges in the industry: unpredictable environment, complexity stemming from the system, scenario, or modeling, data accessibility, and, finally, lack of support standards and guidelines to the adoption of testing. 

Brito et al.~\cite{brito2022integration} promotes integration testing in ROS by releasing a testing framework for checking functional behavior or reaching a high structural coverage thus revealing common faults in mobile robotic systems. Similarly, Laval et al.~\cite{laval2013methodology} introduces a methodology to support the definition of repeatable, reusable, and semi-automated testing for mobile robotic systems. Babi{\'c} et al.~\cite{babic2020vehicle} implements a testing framework targeting heterogeneous marine robots to benefit from simulations while also reflecting the complexity of the real world. Their approach contributes to developing a vehicle-in-the-loop system that can seamlessly be swapped during testing. Biagola and Tonella~\cite{Biagola:TSE24} proposed a test generator for online testing of autonomous driving systems (ADS), namely GENBO (GENerator of BOundary state pairs). GENBO works by mutating the driving conditions of the ego vehicle controlled by the DNN model, within a fixed driving condition scenario from an existing failure-free environment instance. Although the approach is not particularly tailored for ROS-based systems and is limited to the ADS domain, the methodology could shed some light on improving runtime models through exploratory test generation of ROS-based systems. 

{\bf Field-based Testing.} Uniquely, Ortega et al.~\cite{ortega2022testing} report the experience of three developers in performing exploratory and endurance field tests in the Kelo AD disinfection (mobile) robot. Differently from our approach, their work defines a necessary first step to exploring field testing practices in the company's particular context. We, instead, survey the literature and repositories and synthesize guidelines based on data.

Moreover, Bertolino et al.~\cite{bertolino:2021} and  Gazzola et al.~\cite{Gazzola2022} discuss Field Testing as a technique for attaining confidence in software-intensive systems. Bertolino surveys the literature and classifies field-based testing techniques of ex-vivo, offline, and online field testing approaches according to their own taxonomy. Gazzola et al.~\cite{Gazzola2022} presented the concept of field-ready test cases, tests designed to run in production environments. They showcase the usefulness of testing JFreeChart and Apache Commons Lang libraries in the field by exposing faults that escape testing
during development. Differently, our work focuses on understanding field-based testing and runtime verification for robotics systems.

Riedesel~\cite{jamie2021telemetry} defines software telemetry as a set of practices including centralized logging, distributed tracing, and security information event management to handle data from production until delivery to consumers such as software engineers, customer support, or compliance teams. Riedesel's book~\cite{jamie2021telemetry} offers a practical view of telemetry (i.e. monitoring) systems and dives into a pipeline for telemetry as a conceptual architecture for building telemetry systems. Differently from our work, the book does not consider issues that are specific to ROS-based systems and, although their suggestions are fruitful to the discussion at a higher level, they do not help to solve ROS particularities. Moreover, we present guidance targeting software development that should facilitate telemetry (in their terms), which is not discussed in the book. 

{\bf Guidelines.} We also discuss existing studies that propose guidelines to support the development of robotic systems. Related works on specific guidelines can be found in the description of every guideline, as summarized in Tables~\ref{tab:guid_sources1} and \ref{tab:guid_sources2}, especially columns {\em specific search} and {\em exemplars} and in the detailed description of the guidelines, where, for each guideline, we provide some related works and exemplars. The guidelines are described as a whole in a website destined for them~\cite{guidelines}. 


Malavolta et al.~\cite{malavolta2021mining} propose 47 guidelines for supporting roboticists in architecting ROS-based systems. The guidelines are mined from a dataset of 335 GitHub repositories containing open-source ROS-based systems. Our work is also focused on ROS-based systems, but instead of focusing on architectural aspects, we focus on field-based testing and runtime verification.  

Weyns et al.~\cite{Weyns2022} provide a set of guidelines for artifacts that support industry-relevant research on self-adaptation. Artifact providers can use these guidelines for aligning future artifacts with industry needs. The guidelines are based on data obtained from a survey of practitioners and have been defined during working sessions at the SEAMS 2022 (17$^{th}$ International Symposium on Software Engineering for Adaptive and Self-Managing Systems) conference. In~\cite{Weyns:ACSOS21} Weyns et al. present six principles for engineering smarter CPS covering the main stages of the lifetime of CPS: domain engineering, design, operation, and evolution. In their work, they raise the need for both new toolchains and engineers with a deep-rooted understanding of how to develop software able to adapt and evolve under continuous change. None of these guidelines are particularly tailored to robotics, as we do instead; also, we focus on field-based testing and runtime verification.

In the healthcare domain, Bombarda et al.~\cite{BOMBARDA2022} build on their experience in certifying software produced under emergency. By ``under emergency", the author refers to development under strict and pressing time constraints and with the difficulty of establishing a heterogeneous development team made of volunteers. The paper's contribution consists of lessons learned and guidelines, aiming to assist developers in producing safety-critical devices in similar emergencies. The authors discuss each guideline's benefits and risks with the guidelines. These guidelines focus on a specific domain different from robotics and focus on a specific setting, i.e., certifying software produced under emergency.

\section{Conclusion}
\looseness=-1
We proposed 20 guidelines for developers and QA teams that aim at testing and verifying ROS-based robotic systems in the field. Iteratively using design science, we identified and constructed them via a literature review of runtime verification and field-based testing studies on robotic
systems, and mining ROS-based application repositories. 
Grounded on data, the guidelines were deemed actionable by robotics practitioners and researchers who answered a questionnaire that assessed whether the guidelines were clear, useful, and applicable in their work domain. 
Our catalogue of guidelines is extensible and the connected website provides instruments to collect suggestions and recommendations from other researchers and practitioners interested in the domain.   

\looseness=-1
Our guidelines are relevant for researchers and practitioners. The former can benefit from promising future research directions highlighted in our mapping from known challenges to guidelines, e.g., lack of formal specifications, isolation strategies, richer monitoring reactions, and support to imprecise traces. 
The latter can find best practices and recommendations on designing systems to facilitate effective verification and validation of their ROS-based systems at runtime.

\section*{Acknowledgment}

This work is supported by the Wallenberg AI, Autonomous Systems and Software Program (WASP) funded by the Knut and Alice Wallenberg Foundation.
The authors acknowledge the support of the PNRR MUR project VITALITY (ECS00000041), Spoke 2 ASTRA - ``Advanced Space Technologies and Research Alliance", of the PNRR MUR project CHANGES (PE0000020), Spoke 5 ``Science and Technologies for Sustainable Diagnostics of Cultural Heritage'', the PRIN project P2022RSW5W -
RoboChor: Robot Choreography, the PRIN project 2022JKA4SL - HALO: etHical-aware AdjustabLe autOnomous systems, the European Center Agri-BioSERV (SERvices for AGRIfood and BIOmedicine market),
and of the MUR (Italy) Department of Excellence 2023 - 2027 for GSSI. The work of P. Pelliccione was also partially supported by the Centre of EXcellence on Connected, Geo-Localized and Cybersecure Vehicles (EX-Emerge), funded by the Italian Government under CIPE resolution n. 70/2017 (Aug. 7, 2017). This work was financed in part by FAPDF under Call 04/2021 and CNPq process 313215/2021-9.

\bibliographystyle{IEEEtran}
\bibliography{main}

\vskip -1\baselineskip 
\begin{IEEEbiography}[{\includegraphics[width=1in,height=1.25in,clip,keepaspectratio]{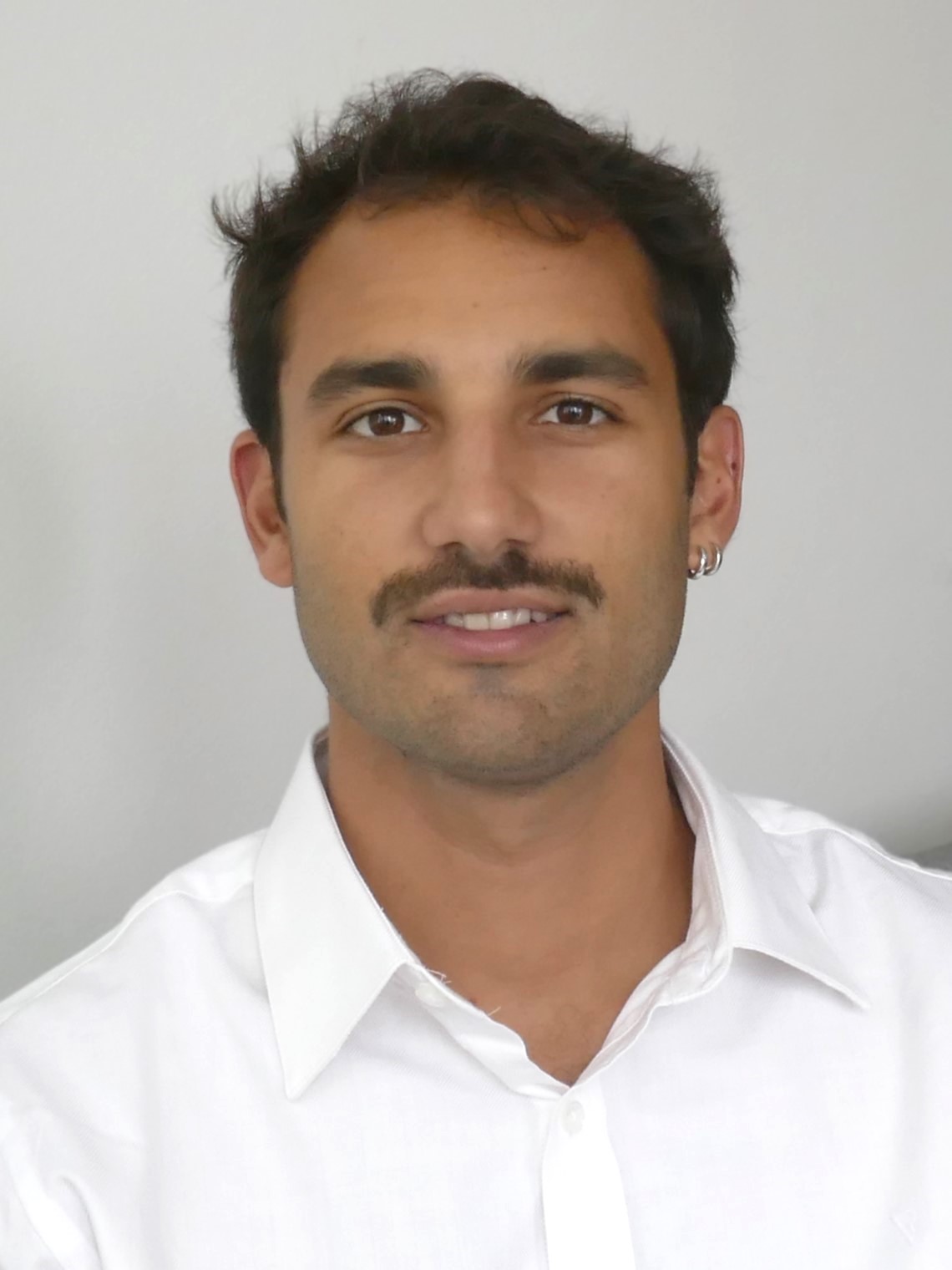}}]{Ricardo Caldas}
is a Ph.D. candidate at the Chalmers University of Technology in Sweden. 
His research topic is software design and assurance provision for resilient autonomous systems. He focuses on resilience as a key enabler to long-living autonomous systems by contributing to software architecture and testing for multi-agent heterogeneous applications, control-based self-adaptation, and automated explanation of violated properties. His contributions extend to software applications implemented in the robot operating system (ROS), in several domains such as robotics, autonomous driving, and healthcare. 
He received his master's degree in Computer Science from the University of Bras\'{i}lia in Brazil in 2019. More information is available at \href{https://ricardocaldas.me}{https://ricardocaldas.me}.
\end{IEEEbiography}
\vskip -2\baselineskip 
\begin{IEEEbiography}[{\includegraphics[width=1in,height=1.25in,clip,keepaspectratio]{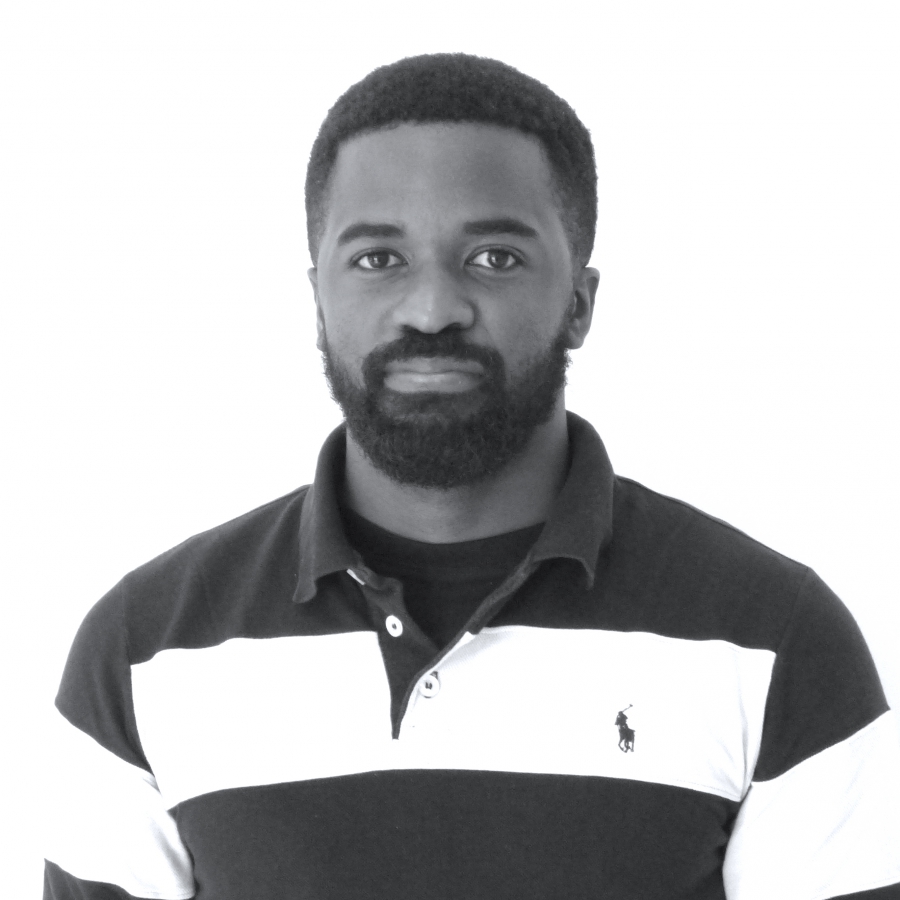}}]{Juan Antonio Pi\~{n}era Garc\'{i}a}
is currently enrolled in the Ph.D. of Computer Science, at the Gran Sasso Science Institute (GSSI), in L'Aquila, Italy. His research topics focus in multi-robot mission coordination and self-adaptation, as well as Imitation Learning techniques based on Deep Learning. Previously, he was a Training Professor at the Computer Engineering Department, Technological University of Havana (Cuba) and a Research Fellow at the Group of Robotics and Mechatronics, Technological University of Havana (Cuba), where he co-led multiple projects to apply robotics to help solve everyday problems and was awarded multiple prizes at international robotics conventions, like the Robotic People Fest, in Colombia.
\end{IEEEbiography}
\vskip -2\baselineskip 
\begin{IEEEbiography}[{\includegraphics[width=1in,height=1.25in,clip,keepaspectratio]{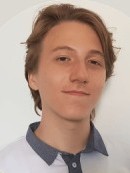}}]{Matei Schiopu}
is a Ph.D. student at the Chalmers University of Technology in Sweden. His research interests include software engineering for cyber-physical and autonomous systems as well as applied machine learning. He received his Master's degree at the Chalmers University of Technology in 2024, working at Volvo Trucks on a machine learning research thesis together with the Battery Management System team.
\end{IEEEbiography}
\vskip -2\baselineskip 
\vfill
\newpage
\begin{IEEEbiography}[{\includegraphics[width=1in,height=1.25in,clip,keepaspectratio]{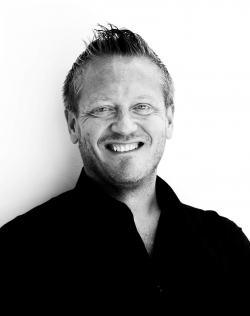}}]{Patrizio Pelliccione}
is a Professor in Computer Science at Gran Sasso Science Institute (GSSI, Italy) and Director of the Computer Science area. Patrizio is also adjunct professor at the University of Bergen, Norway. His research topics are mainly in software engineering, software architecture modeling and verification, autonomous systems, and formal methods. He received his PhD in computer science from the University of L'Aquila (Italy). Thereafter, he worked as a senior researcher at the University of Luxembourg in Luxembourg, then assistant professor at the University of L'Aquila in Italy, then Associate Professor at both Chalmers $\vert$ University of Gothenburg in Sweden and University of L'Aquila.
He has been on the organization and program committees for several top conferences and he is a reviewer for top journals in the software engineering domain. He is very active in European and National projects. In his research activity, he has collaborated with several companies. More information is available at http://patriziopelliccione.com.
\end{IEEEbiography}
\vskip -2\baselineskip 
\begin{IEEEbiography}[{\includegraphics[width=1in,height=1.25in,clip,keepaspectratio]{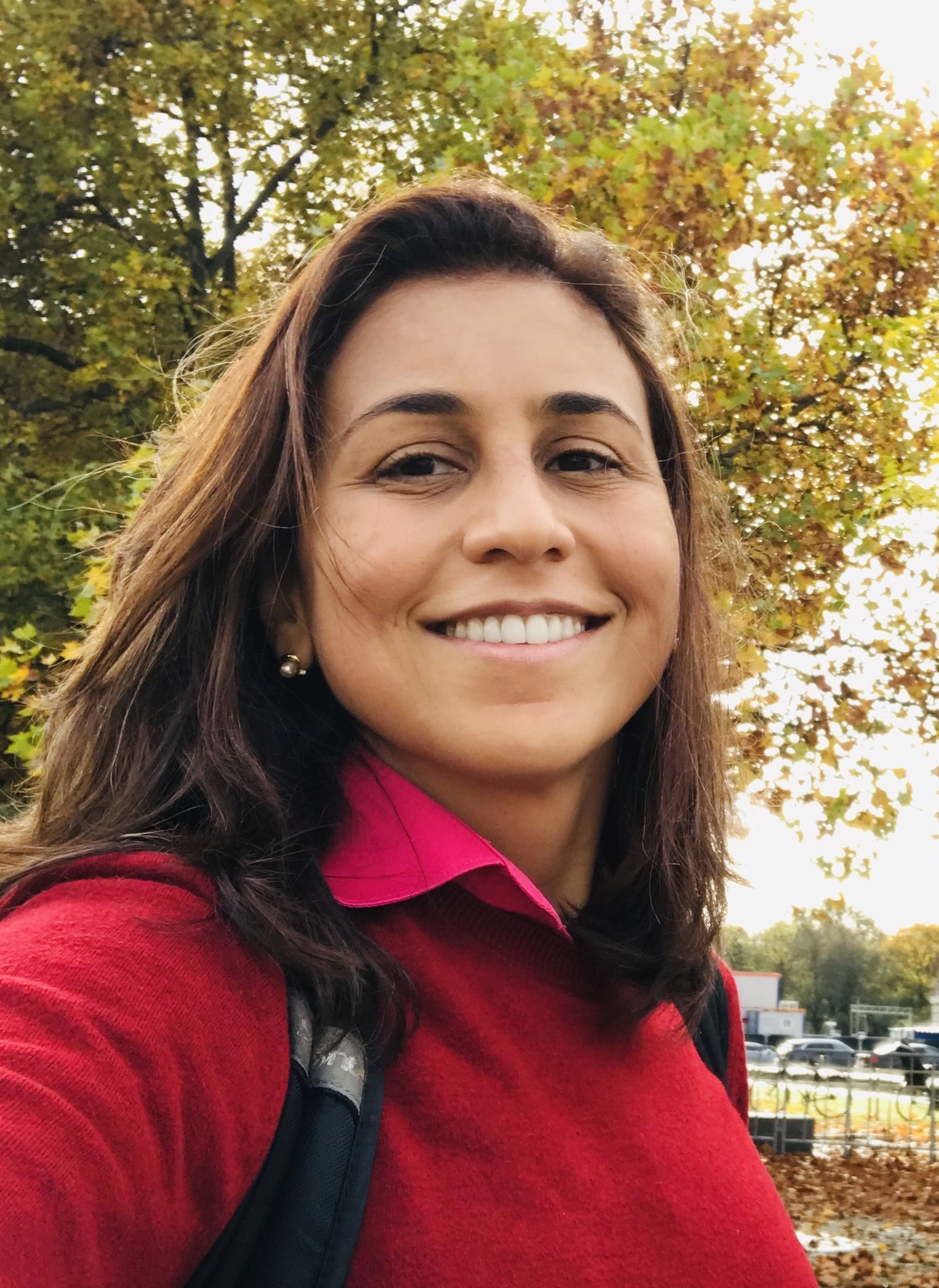}}]{Genaína~Rodrigues}
is an associate professor with the Department of Computer Science at the University of Brasília. Her research interests include the mutual collaboration between smart autonomous systems engineering, in particular robotic systems, and software engineering mainly through goal-oriented requirements engineering and verification both at design and at runtime. She received her Ph.D. in the Computer Science from University College London. In 2020, she was awarded a fellowship as an experienced researcher through CAPES/Alexander von Humboldt Programme to conduct research at the Humboldt Universität zu Berlin in Germany. She has also served highly ranked conferences and journals in Software Engineering worldwide. 
\end{IEEEbiography}
\vskip -2\baselineskip 
\begin{IEEEbiography}[{\includegraphics[width=1in,height=1.25in,clip,keepaspectratio]{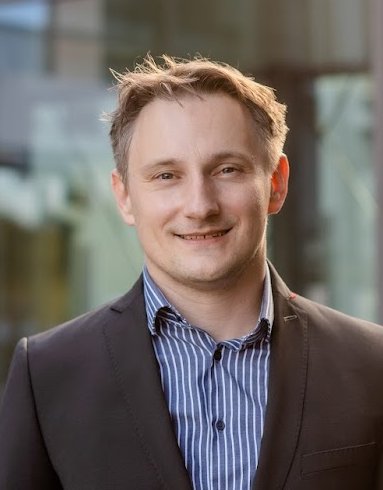}}]{Thorsten Berger}
 is a Professor in Computer Science at Ruhr University Bochum in Germany. His research focuses on automating software engineering for the next generation of intelligent, autonomous, and variant-rich software systems, exploring new ways of software creation, analysis, and evolution. He received the PhD degree in computer science from the University of Leipzig in Germany in 2013. Thereafter, he worked as a Postdoctoral Fellow at the University of Waterloo in Canada and the IT University of Copenhagen in Denmark, and as an Associate Professor at Chalmers\,$\vert$\,University of Gothenburg in Sweden. He received a fellowship from the Royal Swedish Academy of Sciences and the Wallenberg Foundation, one of the highest recognitions for researchers in Sweden. He received two best-paper and two most influential paper awards, as well as his service was recognized with distinguished reviewer awards at A* conferences.
\end{IEEEbiography}
\vfill


\end{document}